\documentclass[]{interact}

\usepackage{adjustbox}
\usepackage{epstopdf}
\usepackage{subfigure}
\usepackage{xcolor}
\usepackage{xr-hyper}
\usepackage{hyperref}
\usepackage{booktabs}
\usepackage{natbib}
\usepackage{caption}
\usepackage{graphicx}
\usepackage{xr}
\usepackage{cleveref}

\bibpunct[, ]{(}{)}{,}{a}{}{,}
\renewcommand\bibfont{\fontsize{10}{12}\selectfont}

\theoremstyle{plain}

\theoremstyle{definition}

\theoremstyle{remark}

\let\realurl\url
\renewcommand{\url}[1]{%
	\realurl{#1}%
	\wlog{URLX #1 }%
}

\makeatletter
\newcommand*{\addFileDependency}[1]{
\typeout{(#1)}
\@addtofilelist{#1}
\IfFileExists{#1}{}{\typeout{No file #1.}}
}\makeatother

\newcommand*{\myexternaldocument}[1]{%
\externaldocument{#1}%
\addFileDependency{#1.tex}%
\addFileDependency{#1.aux}%
}

\myexternaldocument{interacttfvsample_Supplementary}

\begin{document}


\title{A Spatial-statistical model to analyse historical rutting data}


\author{
	\name{N.~O. A. S Jourdain\textsuperscript{a}\thanks{CONTACT N.~O. A. S. Jourdain. Email: natoya.jourdain@ntnu.no}, I. Steinsland\textsuperscript{a}, M. Birkhez-Shami\textsuperscript{a}, E. Vedvik\textsuperscript{c},   W. Olsen\textsuperscript{d}, D. Gryteselv\textsuperscript{b}, D. Siebert\textsuperscript{b} and A. Klein-Paste\textsuperscript{a}}
	\affil{\textsuperscript{a}Norwegian University of Science and Technology (NTNU), Trondheim, Norway; \\ \textsuperscript{b}Statens Vegvesen, Trondheim, Norway; \\
		\textsuperscript{c}Fremtind Forsikring, Trondheim, Norway;
	}
    }

\maketitle

\begin{abstract}
	
	Pavement rutting poses a significant challenge in flexible pavements, necessitating costly asphalt resurfacing. To address this issue comprehensively, we propose an advanced Bayesian hierarchical framework of latent Gaussian models with spatial components. Our model provides a thorough diagnostic analysis, pinpointing areas exhibiting unexpectedly high rutting rates. Incorporating spatial and random components, and important explanatory variables like annual average daily traffic (traffic intensity), asphalt type, rut depth and lane width, our proposed models account for and estimate the influence of these variables on rutting. This approach not only quantifies uncertainties and discerns locations at the highest risk of requiring maintenance, but also uncover spatial dependencies in rutting (millimetre/year). We apply our models to a data set spanning eleven years (2010-2020). Our findings emphasise the systematic unexplained spatial rutting effect, where some of the rutting variability is accounted for by spatial components, asphalt type, in conjunction with traffic intensity, is also found to be the primary driver of rutting. Furthermore, the spatial dependencies uncovered reveal road sections experiencing more than 1 millimeter of rutting beyond annual expectations. This leads to a halving of the expected pavement lifespan in these areas. Our study offers valuable insights, presenting maps indicating expected rutting, and identifying locations with accelerated rutting rates, resulting in a reduction in pavement life expectancy of at least 10 years.

\end{abstract}


\section{Introduction}

Pavement maintenance is essential to ensure road safety and prevent excessive depreciation of roads. Rutting is a major problem in flexible pavement structures and it is one of the major causes why pavements need asphalt resurfacing. Rutting is a distress mechanism that causes longitudinal surface depressions within the wheel path of the asphalt concrete layer. It can substantially affect traffic safety \citep{hao2009rutting} due to impaired steerability and friction reduction due to standing water in the ruts. Ruts occur under the combined actions of traffic loads \citep{zhang2017characterizing} and is affected by environmental conditions, e.g., climate, seepage, temperature, weather \citep{ranadive2016pavement, gudipudi2017impact, zhang2023regional}, and the inherent viscoelastic properties of the construction \citep[e.g., thickness in surface layers or voids in the mixture][]{gu2017characterization, cheng2023predicting}. Rutting may occur at the various layers \citep{lea2015spatial, alhasan2018incorporating} or various stages (e.g., primary, secondary or tertiary) \citep{korkiala2007relating, al2009three, gupta2014critical} of a constructed pavement. In cold climates, rutting often determines the longevity of the top layer of flexible pavements \citep{lundy1992wheel}, and the dominating mechanism is wear from studded tires \citep{snilsberg2008pavement}, as opposed to permanent deformations in the mastic.

The rut depth can be effectively measured by laser profilometry \citep{thodesen2012review}. Many road agencies or road maintenance contractors map their road network regularly as input for their Pavement Management System (PMS). In Norway, all state roads are scanned every year, and this measurement program have been providing pavement condition data into a single database (\texttt{ROSITA}) since 2010. The Norwegian PMS system can provide automatically generated reports, for example which sections exceed the maximum allowed rut depth defined in the maintenance standards. However, deeper diagnostic analysis to identify areas that exhibit unexpectedly high rutting rates still requires manual labor. Such type of data analysis is typically only initiated based on suspicion and are very time consuming.

Acknowledging that the value of long time series of pavement condition data has not been fully utilized, the Norwegian Public Roads Administration initiated a research project that aims for more automated pavement diagnostics. Our working hypothesis is that advanced statistical analysis of the available data can provide a better understanding of how the pavement condition of a whole network develops and allow automated identification of segments that perform better or worse than expected. Moving from “manual, suspicion-based” diagnostics towards automated identification of problem areas enables pavement managers to concentrate their diagnostic resources on finding the correct countermeasures, rather than finding the problem areas.  Identifying sections with unexpectedly high rutting levels requires that one compares with the ‘expected performance' in terms of rutting. 

Several empirical models have been proposed in the past to predict the performance of pavements \citep{hu2022review, yao2023effects, yu2023prediction} and these models typically include rutting. Empirical models usually relate the accumulation of distresses (e.g., rut, roughness, cracks) in the pavement layers to the load repetitions (annual average daily traffic) and other affecting factors (e.g., asphalt type, weather, width of the road, layer thickness). Empirical models are known to suffer from limited applicability \citep{gupta2014critical}, so when the conditions (for example road construction, weather conditions or damage mechanisms) during the data collection differs from the conditions where the model is applied, one cannot expect to get reasonable results. Empirical models developed for pavement wear in cold climates exists \citep{jacobson2005prediction}, but the challenge of limited predictability remains present. Calculating the expected rutting for sections based on an empirical model is therefore likely to give large uncertainty.

Another modelling framework for predicting pavement performance from local historical data has been based on foundation layers and stages of deterioration and models of transition probability matrices (TPMs) \citep{chou2008pavement, perez2019rigid, yamany2020prediction}. This has some weaknesses, including its usability for the prediction of conditions based on a limited number of discrete classes \citep{vermunt1999discrete, pulugurta2009pavement, standfield2014markov} increasing imprecision and inaccuracy when assigning TPMs to the various deterioration stages; limitation in their generalizability to other flexible pavement data; and their lack of potential to account for spatial heterogeneity in the data \citep{saba2006performance, chen2016sigmoidal}

Spatial models can provide additional explanation by accounting for trends in location, i.e., along the road. In the case of pavements, it is likely that sections close to each other behave more similar than sections that are far apart.  These location components cannot explain the cause of the behaviour but they do have explanatory value and can be utilized to identify ‘hotspots', predict future rut depths and quantify uncertainties. 

Spatial models have been proposed for understanding pavement behaviour earlier \citep{lea2015spatial, alhasan2018incorporating, ebrahimi2019estimation}. \cite{lea2015spatial} considered the applicability of spatial variability models on a test pavement containing a modified binder constructed for heavy vehicle simulator testing and found that the implementation of the {spatial statistical} techniques gave good insights concerning pavement variability following the complexity of pavement construction. \citep{alhasan2018incorporating}, on the other hand, extended the more traditional mechanistic-empirical pavement prediction modelling framework by incorporating spatial variability and uncertainty in pavement foundation layers. \citep{ebrahimi2019estimation} estimated asphalt surfacing lifetime for open road infrastructure on Norway road network (excluding bridges and tunnels) in Troms, S\o r-Tr\o ndelag and Vest-Agder. Their analysis showed that narrower road width was noted to result in shorter pavement lifetimes, particularly in Troms, where studded tires are used one month earlier than other locations, e.g., Vestland or Tr\o ndelag in Norway.

\subsection{Objectives}
\label{sec:ojectives_1}

Our primary goal is to develop spatial-statistical models that can pinpoint pavement locations exhibiting unexpected and pronounced rutting. To achieve this, we use eleven years (2010-2020) of pavement rutting data from the European highway route E14, stretching $67.1$ kilometres from Stj\o rdal, Norway to Storlien on the Swedish border. This data, collected annually, provides a robust historical record that is essential for identifying trends and anomalies in rutting patterns. By utilising this extensive dataset, we aim to enhance our understanding of rutting dynamics over time and across different sections of the road.

To analyze the rutting phenomenon, we develop a hierarchical Bayesian framework utilizing latent Gaussian models (LGMs) \citep{Jourdain2024}. These models are particularly suited for this type of analysis due to their flexibility and ability to explicitly account for spatial and random components \citep[e.g.,][]{dey2000generalized, chu2005gaussian, banerjee2008gaussian, Rue2009}. The LGMs enable us to capture the underlying processes governing rutting and provide probabilistic predictions, which are crucial for understanding and managing pavement performance. Advanced statistical analysis is implemented through the Integrated Nested Laplace Approximation approach (\texttt{INLA}) for performing Bayesian inference. This approach is well-suited for LGMs and will allow us to efficiently handle the complex computations involved in our models. By using \texttt{INLA}, we aim to achieve high computational efficiency and accuracy in our analysis.

Incorporating a selection of explanatory variables known to influence pavement rutting is another key objective. These variables include asphalt type, Average Annual Daily Traffic (AADT), rut depth from the previous year, and lane width. The inclusion of these factors is based on their established impact on rutting \citep{chan2010investigating,  lang2012comparison, thodesen2012review, ebrahimi2019estimation, yan2020analysis, zhao2020factors,pan2021rutting, li2023design}. By integrating these variables into our model, we can better account for the diverse factors that contribute to rutting, thereby improving the accuracy and reliability of our estimations.

Identifying areas with unexplained or accelerated rutting is a crucial part of our objectives. By analyzing the data with our models, we pinpoint sections of the road that exhibit unexpected rutting patterns.

Therefore, we provide practical insights for road authorities to use the model results for better pavement maintenance planning. By shifting from manual, suspicion-based diagnostics to a more efficient, automated approach, we hope to enable pavement managers to concentrate their diagnostic resources on identifying the correct countermeasures rather than merely locating problem areas. This will ultimately lead to more effective maintenance strategies and improved road safety.

\section{Materials} 
\label{sec:methodsandmaterials}

\subsection{Area of study}
\label{sec:areaofstudy}
Our case study area is the European route E14 in Trøndelag County, Norway, stretching from Stjørdal to Storlien on the Norwegian-Swedish border. This major highway, managed by the Norwegian Public Roads Administration, consists of flexible bituminous mixtures. Located at approximately $63^{\circ}  \ 19^{\prime} \ 1.80^{''} \ \mathrm{N}$ latitude and $14^{\circ}  \ 11^ {\prime} \ 24.00^{''} \ \mathrm{E}$ longitude, the road spans 67.1 km. 

Most of the road has two lanes, but within approximately 2.5 km from Stjørdal, it expands to four lanes. Two-lane stretches are around 6 meters wide, while four-lane sections are up to 13 meters wide. The expected lifetime of the road is 20-30 years after construction. 

The climate is subarctic or cold-temperate, with temperatures varying by season: winter (-2°C to 2°C), spring (5°C to 12°C), summer (15°C to 22°C), and autumn (10°C to 15°C) \citep{sygna2001virkninger,institutt2010ekstremvarsel}. On average, air surface temperatures are at most 18 °C (see Figure \ref{fig:temperature_APPENDIX} in Appendix \ref{sec:temperature_appendix})  .

For analysis, we focus on one lane (lane 1) in the direction from Stjørdal to Storlien.

\subsection{Data}
\label{sec:Data}

The data for this research is sourced from the Norwegian Road Database (NVDB) and the \texttt{Rosita} database. NVDB, primarily developed by the Norwegian Public Roads Administration (NPRA), includes 93,500 kilometers of public roads, comprising national, county, and municipal roads \citep{stephansen2019}. The Rosita database provides extensive and calculated data from measuring systems (see Section \ref{sec:rutdepthandrutting}).

We use eleven years (2010-2020) of pavement rutting data from the European highway route E14, covering a 67.1-kilometer stretch from Stjørdal, Norway, to Storlien on the Swedish border. Collected annually, this data set offers a comprehensive historical record crucial for detecting trends and anomalies in rutting patterns.

Explanatory variables for expected rutting include annual average daily traffic (AADT), asphalt type, rut depth from the previous year, and lane width. Table \ref{tab:data2018} summarizes the data used, and Figure \ref{fig:mapofnorway_data_1} and Supplementary Material \ref{supplementary:datasummary} provide additional details.

\renewcommand{\arraystretch}{1.2}
\begin{center}
	\begin{table}[!htbp]
		\small
		\caption{Nomenclature and Variables}
		\begin{tabular}{ll} 
			\hline \\[1.0ex] 
			\textbf{Terms and Variables} & \textbf{Description/Value} \\[1.0ex]
			\hline \\ 
			Rut depth ($d_{i,t}$, mm) & Measurement for year $t$ for all $20$-metre road segments ($i$); mean $=12.11$, sd $=6.78$ \\ [0.5ex]
			Lane width (${w_{i}}$, m)  & Half (or quarter) of width of 2-lane (or $2-2$ lane) used; mean  $=2.99$, sd $=0.18$ \\[0.5ex] 
			Asphalt type & Asphalt concrete (Ac), asphalt gravel concrete (Agc), stone mastic asphalt (Sma) \\[0.5ex] 
			AADT (vehicles/day) & Range: 568-18,500, adjusted for the number of lanes \\[0.5ex] 
			$r_{i,t}$ (mm/year) & Rutting, calculated as the change in rut depth ($d_{i,t}$) between years $t$ and $t-1$ \\[0.5ex]
			\hline 
		\end{tabular}
		\label{tab:data2018}
	\end{table}
\end{center}

\subsubsection{Rut Depth and Rutting}
\label{sec:rutdepthandrutting}

Rut depth is measured annually using the \texttt{ViaPPS} (Pavement Profile Scanner - laser scanner), developed by \texttt{ViaTech} in collaboration with the NPRA \citep{peraka2020pavement}. Measurement positions are determined with high-accuracy GPS (CPOS/DPOS --- satellite-based position service with centimetre/decimetre accuracy), ensuring precise location data. The rut depth is estimated by comparing the height of the ridges formed between the wheel paths, which involves identifying the two lowest points in the wheel path and measuring the distance to the top of the ridge. Additionally, the laser covers an approximate 4-meter footprint, ensuring that the vehicle remains within the wheel tracks and avoids crossing the centerline or shoulder. Thus, minor deviations (10-50 cm) do not affect the calculations. Annual rut depths along the road range from 0 to 41 mm, measured at 20-metre intervals. There are 3355 road segments measured for annual rut depth.

Rutting, denoted as $r_{i,t}$ (mm/year), is calculated as the difference in rut depth between years, as shown in Equation \ref{eqn:rutitngdefinition}. We denote  the rut depth variable at road segment  $i=1,...,n$ in year $t$ as  $d_{i,t}$, and $r_{i,t}$ as \emph{rutting} (mm/year). Rutting for year $t= 1,...,T$ at segment $i$ is defined as
\begin{center}
	\begin{equation}
	r_{i,t} = d_{i,t} - d_{i, t-1}.
	\label{eqn:rutitngdefinition}
	\end{equation}
\end{center}
When repavement occurs, rut depth decreases, sometimes resulting in large negative values of the rutting ($r_{i,t}$). Following \cite{vedvik2021spatial}, large negative rutting values are considered missing and set to NaN (not a number):

\[ r_{i,t} =
\begin{cases} 
r_{i,t} & \text{if } r_{i,t} \ge -\frac{d_{i,t}}{2} \\ 
\text{NaN} & \text{otherwise,} 
\end{cases}
\]
and the small negative values kept in the dataset are considered as measurement uncertainties and are accounted for in the random variability. Some missing rutting data may occur due to measurement uncertainty or data collection process. For available observations, the overall mean rutting and segment mean rutting (over all years) are calculated as $	\bar{r} = {1}/{n}\sum_{\forall t} \sum_{\forall i} r_{i,t}$ and $	\bar{r}_{i} = {1}/{T} \sum_{t=1}^{T} r_{i,t}$, respectively.

\subsubsection{Lane Width}
\label{sec:lanewidth}
In the database, road width and the number of lanes are available. The first 2.5 km from Stj\o rdal has four (4) lanes, the remaining stretch has two lanes. Lane width is calculated by dividing the road by the number of lanes.

\subsubsection{Annual Average Daily Traffic (AADT)}
\label{sec:annualaveragedailytraffic}

\begin{figure}[!htbp]
	\centering    
	\subfigure[AADT/Asphalt type and lane width]{ 
		\includegraphics[width=0.47\textwidth]{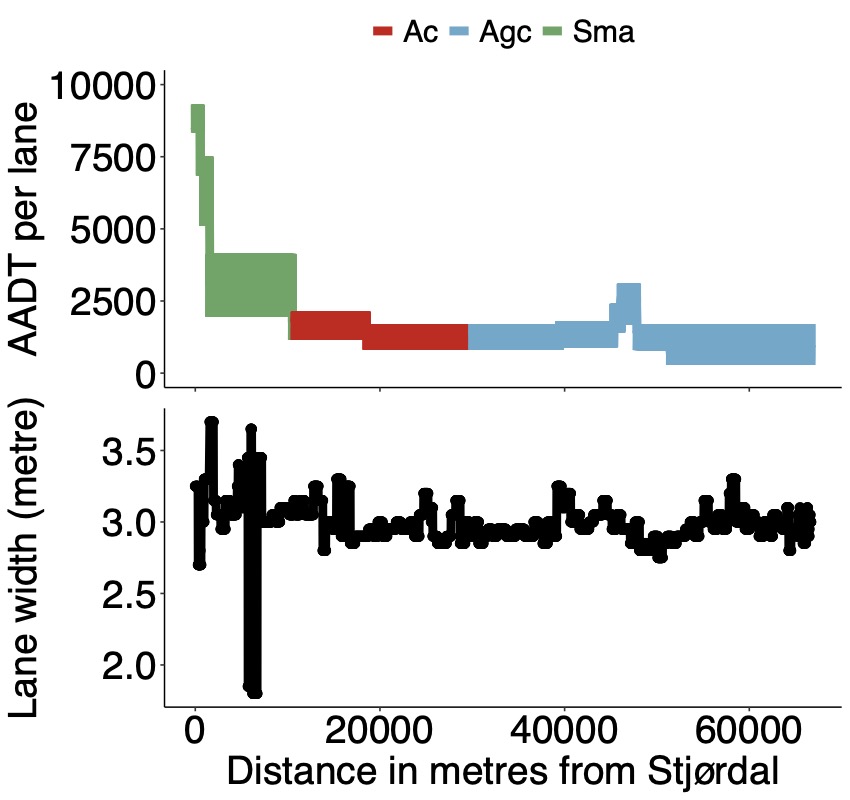}}
	\subfigure[Rutting]{ 
		\includegraphics[width=0.47\textwidth]{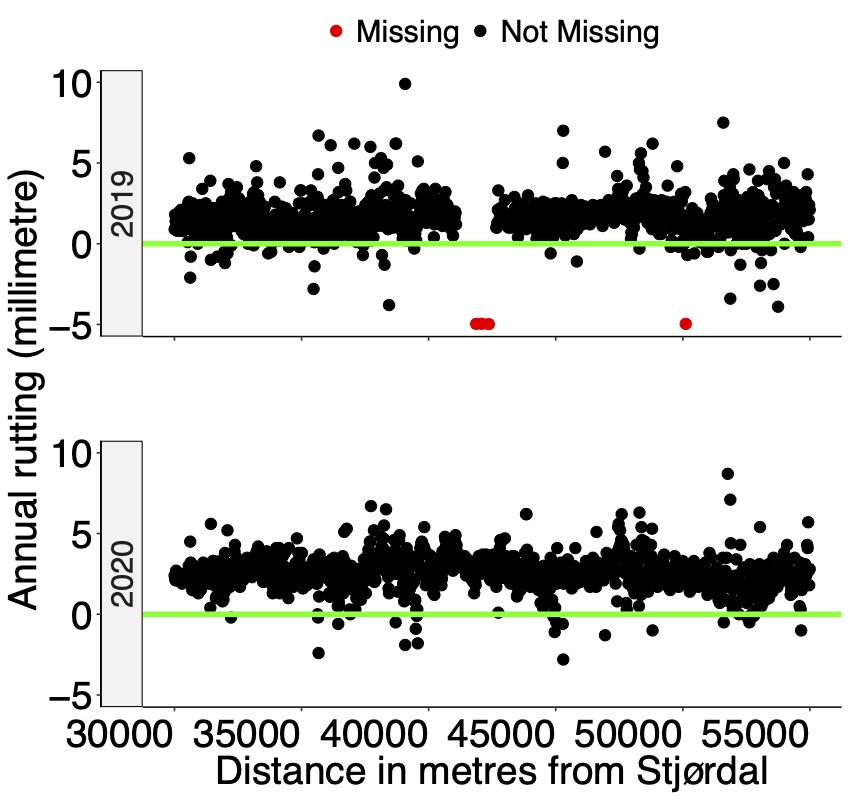}} 
	
	\caption{AADT together with asphalt type, and lane width (a), and rutting for a short stretch of E14 (for years 2019-2020, mm/year --- missing in red) (b).} 
	\label{fig:mapofnorway_data_1}
\end{figure}

The Annual Average Daily Traffic (AADT) for each road segment between 2010 and 2020 varied from 568 to 18,500 vehicles per day (Figure \ref{fig:mapofnorway_data_1}, graph (a)). Fluctuations in traffic volume over short distances are influenced by various factors. Notably, there are considerably larger traffic volumes within Stjørdal to approximately 2.5 km towards Storlien. The European route E14 runs through the city center of Stjørdal, contributing to high traffic levels in the initial few kilometers.

The Norwegian Public Roads Administration (NPRA) recommends maintenance based on AADT values. High-traffic roads should be treated when rut depths exceed 20 mm, while low-traffic roads should be treated when rut depths exceed 25 mm. Adjusting AADT values based on road width is essential. Throughout the analyses, AADT values are expressed in ten-thousands.

\subsubsection{Asphalt Type}
\label{sec:asphalttype}

Asphalt mixtures in cold-temperature regions like Norway are specifically designed to withstand harsh climatic conditions. These mixtures may include Polymer Modified Asphalt (PMA) to prevent cracking due to freeze-thaw cycles and low temperatures, and low-temperature performance grade binders to maintain flexibility at sub-zero temperatures. Dense-graded aggregates are used to minimize water infiltration, crucial for preventing freeze-thaw damage. The primary focus is on enhancing flexibility, low-temperature performance, and resistance to moisture, employing softer binders, dense-graded aggregates, and anti-stripping agents \citep{snilsberg2006analysis, d2008warm, saba2009development, odegaard2015effects}.

Asphalt concrete (\texttt{Ac}), asphalt gravel concrete (\texttt{Agc}), and stone mastic asphalt (\texttt{Sma}) are the primary surface mixtures used in Norway. \texttt{Sma}, known for its superior wear resistance and deformation resistance due to the larger- and high-quality-stone in its composition, is typically employed on high-traffic roads. In contrast, \texttt{Ac} and \texttt{Agc} are commonly used on low-traffic roads, courtyards, sidewalks, and bicycle paths.


\subsection{Exploratory analysis of rutting}
\label{sec:exploratoryanalysis}
\vspace{0.2cm}

\begin{figure}[!htbp]
	\centering    
	\begin{minipage}{\textwidth}
		\subfigure[Observed rutting]{
			\includegraphics[width=0.42\textwidth]{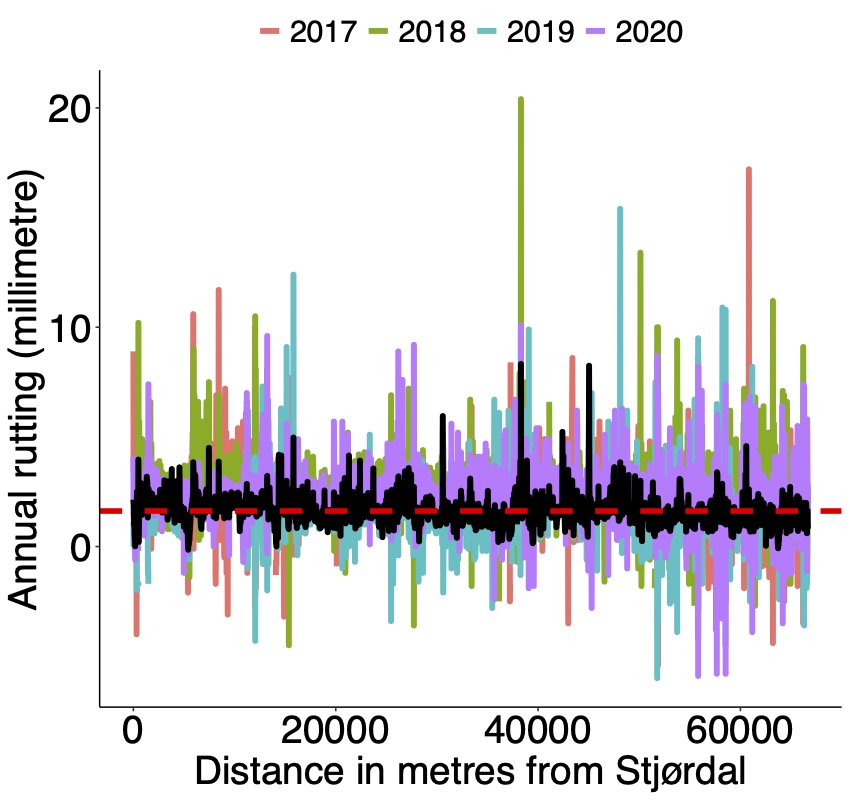}}  
		\hspace{1cm}
		\hfill
		\subfigure[Density plot of rutting]{s
			\includegraphics[width=0.42\textwidth]{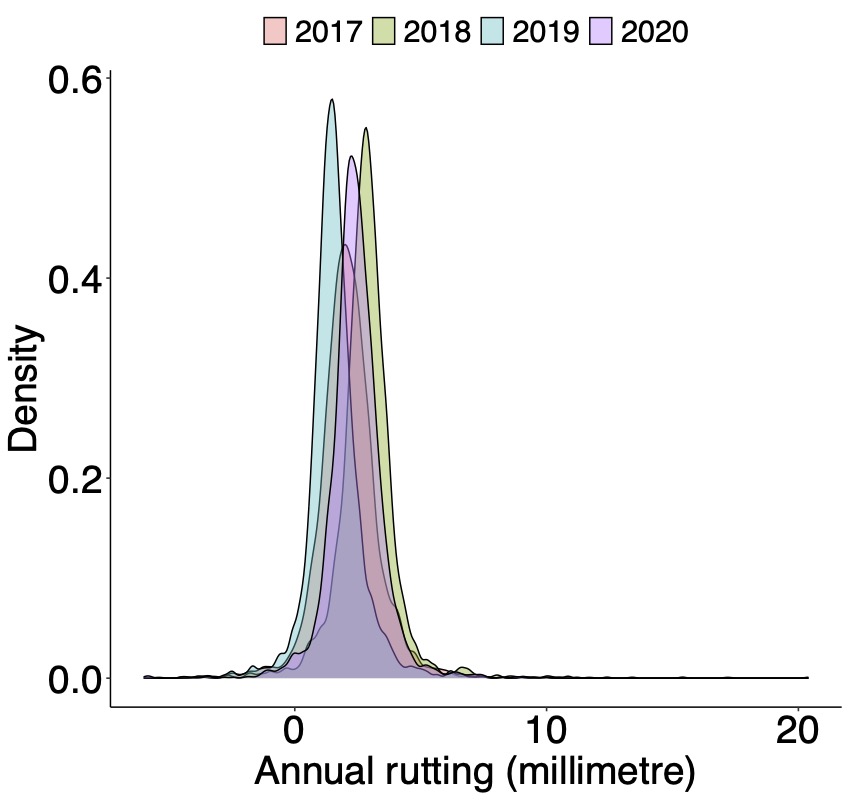}}
		
		\subfigure[Autocorrelation of segment mean rutting]{
			\includegraphics[width=0.42\textwidth]{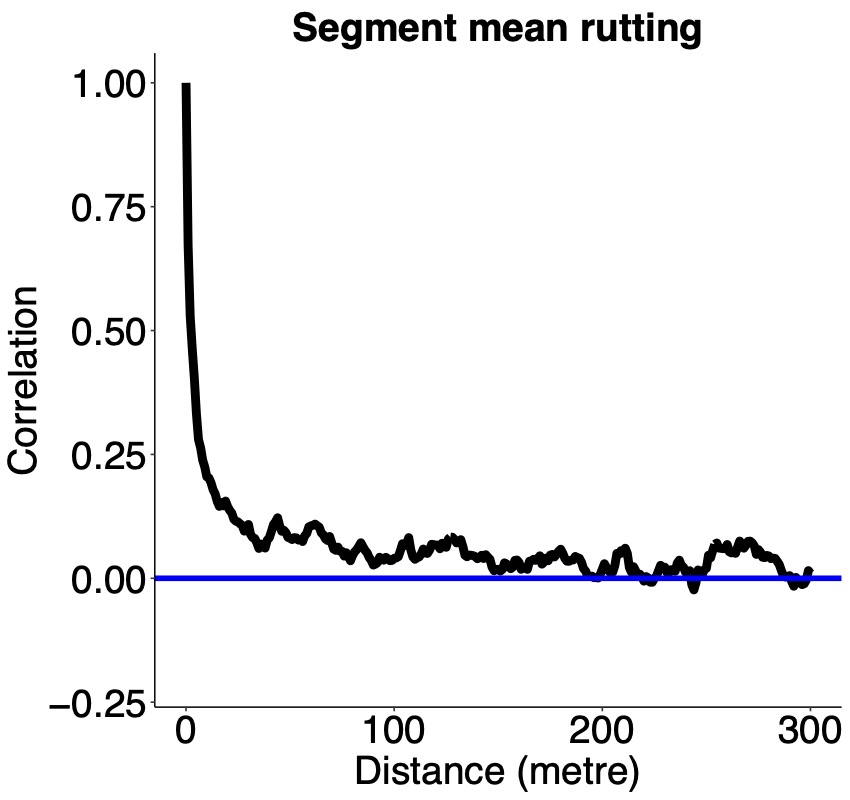}}
		\hfill
		\subfigure[Autocorrelation for years 2011-2012]{
			\includegraphics[width=0.42\textwidth]{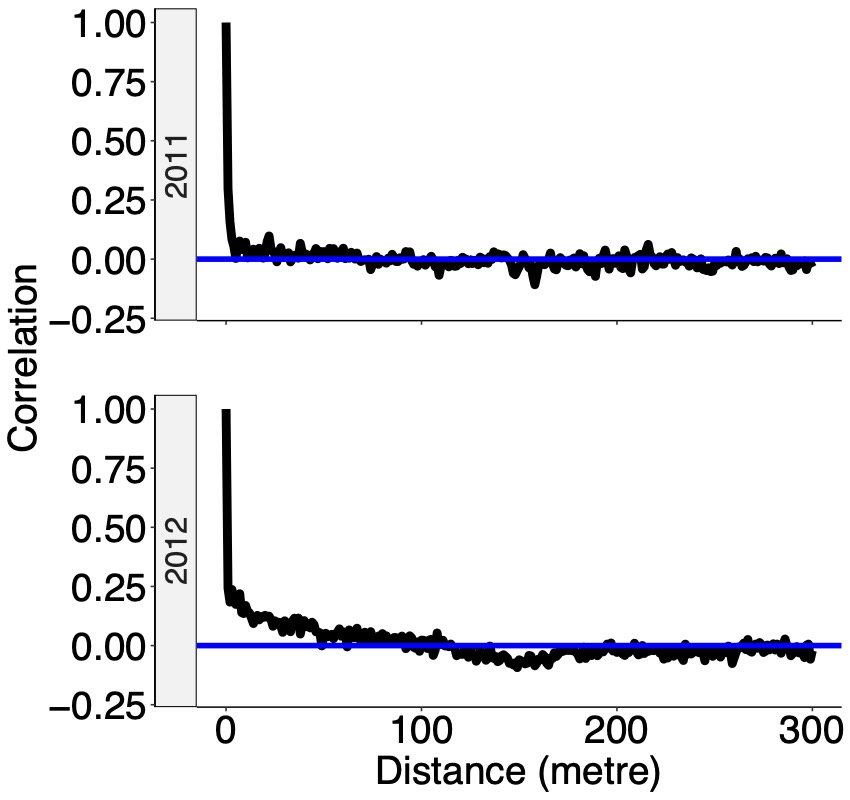}}
	\end{minipage}
	\caption{Graph (a): Observed rutting for each year (2017-2020), the overall mean rutting ($\bar{r}=1.62$ mm/year (for all years, 2011-2020) (\textcolor{red}{\rule[.5ex]{0.4em}{2pt}} \textcolor{red}{\rule[.5ex]{0.4em}{2pt}} \textcolor{red}{\rule[.5ex]{0.4em}{2pt}}), and segment mean rutting ($\bar{r}_{i}$, over all 11 years) (\textcolor{black}{\rule[.5ex]{1em}{2pt}}). Graph (b): Density plots for rutting. Graphs (c) and (d) show how rutting values are correlated with distance. Spatial dependencies in rutting exist up to 150 metres.}
	\label{fig:annualrutting_density_variograms}
\end{figure}

On average, over all stretches and years, the overall rutting is $1.62$ mm/year; see Figure  $~\ref{fig:annualrutting_density_variograms}$ (a). The density plots, Figure $~\ref{fig:annualrutting_density_variograms}$ graph (b), show rutting for all segments for four selected years (2017-2020). Some years having higher rutting levels and are more widely distributed than others.  The lines in graphs (c) and (d) represent the empirical correlation of rutting values with respect to distance. Specifically, graph (c) shows the segment mean rutting over all years (see Section \ref{sec:rutdepthandrutting}), and graph (d) displays the correlation for two selected years (2011-2012). These values are computed from the observed rutting at 20-meter segments, illustrating the empirical correlation as a function of the distance between the segments \citep{szekely2007measuring, jentsch2020empirical}. Based on the data, we present results for distances up to 300 metres.

In graph (d), the x-axis represents the distance in meters, while the y-axis shows the calculated correlation between rutting values. This graph highlights the spatial dependencies of pavement rutting along the road for specific years. When compared to graph (c), which reveals long-term spatial patterns and trends, graph (d) demonstrates that the correlation of rutting values tends to diminish more quickly over shorter distances.

Graph (c) indicates that there are meaningful correlations in rutting values for distances up to approximately 150 meters, suggesting that these measurements capture underlying trends that persist over longer stretches of the road. In contrast, graph (d) shows that for specific years, the correlation between rutting values is important only up to about 100 meters. This shorter correlation distance implies that the rutting patterns observed in a given year are influenced by more localized factors and exhibit greater variability over shorter distances.

The difference in correlation distances between the two graphs shows the importance of considering both short-term and long-term spatial dependencies when analyzing pavement conditions. While long-term data provide a comprehensive view of how road conditions evolve, short-term data, as depicted in graph (d), highlight the influence of yearly variations and localized factors on pavement rutting. This dual approach enables a better understanding of the factors affecting road quality and can inform more effective maintenance and rehabilitation strategies.\\


\clearpage

\vspace{-0.7cm}

\section{Latent Gaussian models for Rutting}
\label{sec:latentgaussianmodels}

A Bayesian latent Gaussian model is a  three-stage hierarchical model consisting of  a likelihood, a latent Gaussian field, and hyperparameters. In this paper, the modelled rutting (i.e., all rutting behavior) at location $s_{i}$ in year $t$ is denoted by $\eta_{t} (s_{i})$ and it is modelled as a latent Gaussian field that includes explanatory variables and spatial effects. We specify six (6) models with different explanatory variables included and with and without spatial and random effects. 

\subsection{Likelihood: observed rutting given modelled rutting}
\label{sec:likelihood}
We denote the observed rutting (detected and measured) for segment $i$ in year $t$ as $r_{i,t}$. The likelihood of the rutting, $r_{i,t}$, at road segment $i=1,...,n$ in year $t = 1,..., T$ is expressed as
\begin{center}
	\begin{equation}
	\displaystyle r_{i,t} \ | \ \eta_{t} (s_{i}) \sim \mathrm{N}\left(\eta_{t} (s_{i}),  \sigma^{2}_{\varepsilon} \right).
	\label{eqn:likelihood_rutting_model_definition_1}
	\end{equation}
\end{center}
Here, $\sigma^2_{\varepsilon}$ is the uncertainty of the Gaussian observations. The unstructured error terms $\varepsilon_{i,t} \stackrel{i.i.d}{\sim} \mathrm{Normal}(0, \sigma^{2}_{\varepsilon})$  are independent and identically distributed (i.i.d).

\subsection{Latent field: modelled rutting}
\label{sec:latenfield}

We model the observed rutting as Gaussian with expected value $\eta_{t}(s_{i})$ and variance $\sigma^{2}_{\varepsilon}$. Further, $r_{i,t}$ is conditionally independent, given the latent field $\eta_{t}(s_{i})$. We only consider models  that have an interaction with AADT, and include a random yearly effect. It is known that  more traffic gives more rutting, so we include traffic (AADT) in the model as an interaction with each of the explanatory variables. This means that for asphalt type, we assume that the rutting is linear for all asphalt types (double traffic means double rutting), but that the different asphalt types can have different rates of deterioration. From the explanatory analysis (Figure \ref{fig:annualrutting_density_variograms}), we found that the rutting seems to be on different levels for different years, and this might be explained by different weather conditions. To account for this, we include a random yearly effect in the model.

\subsubsection{Non-spatial rutting model: based on explanatory variables and year effect}

The model without spatial effects (non-spatial) has the linear predictor
\vspace{-0.4cm}
\begin{center}
	\begin{equation}
	\centering
	\begin{split}
	\eta_{t}(s) & = \beta_{\mathrm{Ac}} z_{\mathrm{Ac}(s,t) }   + \beta_{\mathrm{Agc}} z_{\mathrm{Agc}(s,t) }  + \beta_{\mathrm{Sma}} z_{\mathrm{Sma}(s,t) }  + \beta_{\mathrm{d-1}} z_{\mathrm{d-1}(s,t) } + \beta_{\mathrm{w}} z_{\mathrm{w}(s,t) }  + \gamma_{t}. 
	\label{eqn:model8}
	\end{split}
	\end{equation}
\end{center}
The model assumes that the level of expected rutting (mm/year) is influenced by an interaction effect of the annual average daily traffic (AADT) and some explanatory variables, including the level of rut depth from the previous year ($z_{\mathrm{d-1}}$), the lane width ($z_{\mathrm{w}}$) and asphalt type. This assumption stems from our prior understanding: when there's no traffic, there's no rutting (or rut depth). This insight is grounded in research, such as the work of Hao et al. (2009), which identified increased loading time (traffic loads) as the primary factor accelerating rutting in asphalt pavements.

Let  $z_{\mathrm{AADT} (s,t)}$ denote AADT for location $s$ in year $t$, and  $I_{\mathrm{Ac}(s,t)}$ be an indicator variable that has a value of 1 if location $s$ in year $t$ has asphalt type Ac and $0$ otherwise. Then, $z_{\mathrm{Ac} (s,t)} =I_{\mathrm{Ac}(s,t)} \cdot z_{\mathrm{AADT} (s,t)}$ and, hence, equals AADT if the asphalt type is Ac (asphalt concrete) and $0$ otherwise. Similarly, $z_{\mathrm{Agc} (s,t)}$ and $z_{\mathrm{Sma} (s,t)}$ equal the AADT for location $s$ in year $t$ if the asphalt type is Agc (asphalt gravel concrete) and Sma (stone mastic asphalt), respectively, and $0$ otherwise. Further, $z_{\mathrm{d-1} (s,t)} = d (s,t) \cdot z_{\mathrm{AADT} (s,t)}$ and $z_{\mathrm{w (s,t)}} = \mathrm{w}{ (s,t)} \cdot z_{\mathrm{AADT} (s,t)}$. 

The random yearly effect is assumed to be distributed as $\displaystyle \boldsymbol{\gamma} = (\gamma_{1}, \gamma_{2},...,\gamma_{T})  {\sim}  \mathrm{Normal} (\boldsymbol{0}_{T},\  \tau_{\gamma}^{-1})$.

\subsubsection{Models including spatial effects}
\vspace{0.2cm}

Our model (Equation (\ref{eqn:model8})) does not include several variables known to influence rutting, such as weather, bearing capacity, and proportion of heavy traffic load. These variables are spatially dependent, with segments closer in space likely to be more similar than those further apart. To account for this spatial dependency, we incorporate a Gaussian process (GP) as a spatial random effect along the road. GPs are powerful tools for modeling spatial dependencies, capturing variability due to both known factors (e.g., road characteristics, traffic patterns, and other explanatory variables) and unknown factors not explicitly accounted for in the model. The unknown factors could include unmeasured road characteristics, environmental variables like temperature, or other spatially dependent factors influencing rutting. Our model aims to capture both consistent spatial effects that remain stable over time and spatial effects that vary between years.

Consistent spatial effects, such as road characteristics (e.g., bearing capacity, side slope height, surface and base curvature, and proportion of heavy traffic), are likely to influence rutting consistently over time. Additionally, there are spatial effects that vary between years, such as heavy rain events, temperature variations, or fluctuations in heavy traffic due to construction work. To account for these variations, we include both an annually varying spatial process, $\xi_{t} (\boldsymbol{s})$, for each year $t=1,...,T$, and a common spatial process, $\omega(\boldsymbol{s})$, referred to as the spatial rutting component.

The annually varying spatial process, $\xi_{t} (\boldsymbol{s})$, captures the spatial variability in rutting that changes from year to year, accounting for spatial variations specific to each year, such as the impact of specific weather events or construction activities.

The common spatial process, $\omega(\boldsymbol{s})$, which we refer to as the spatial rutting component, captures the spatial variability in rutting that remains consistent across all years. This includes spatial effects not explicitly accounted for by other variables in the model, such as road characteristics and other spatially dependent factors. This contribution can be either positive or negative.  The linear predictor for the model including spatial effects is described as

\vspace{-0.4cm}
\begin{center}
	\begin{equation}
	\centering
	\begin{aligned}
	\eta_{t}(s) & = \beta_{\mathrm{Ac}} z_{\mathrm{Ac}(s,t) }   + \beta_{\mathrm{Agc}} z_{\mathrm{Agc}(s,t) }  + \beta_{\mathrm{Sma}} z_{\mathrm{Sma}(s,t) }  + \beta_{\mathrm{d-1}} z_{\mathrm{d-1}(s,t) } + \beta_{\mathrm{w}} z_{\mathrm{w}(s,t) }   + \\ & \gamma_{t} +  \xi_{t} (s) + \omega(s),  
	\label{eqn:mode4}
	\end{aligned}
	\end{equation}
\end{center}
\vspace{0.2cm}
where the remaining model parameters are defined as that in Equation \ref{eqn:model8}.

The common spatial process $\omega(\boldsymbol{s})$ in Equations (\ref{eqn:mode4}) is the time constant variable. It is a Gaussian random process $\left\{\omega (\boldsymbol{s}) : \boldsymbol{s} \in \mathcal{D} \subseteq \mathbb{R}^{d} \right\} $ such that for any $n\ge 1$ and for each set of spatial locations $\left(s_{1},...,s_{n}  \right)$ satisfies
\vspace{-0.4cm}
\begin{center}
	\begin{equation}
	\omega(\boldsymbol{s}) = \left\{{\omega} (s_{1}),..., {\omega} (s_{n}) \right\} = \left({\omega}_{1},..., {\omega}_{n}     \right) \sim \mathrm{Normal} \left(\boldsymbol{\mu}^{\omega}, \Sigma^{\omega} \right),
	\label{eqn:commonspatialfield}
	\end{equation}
\end{center}
where $\boldsymbol{\mu}^{\omega} = \left\{\mu^{\omega} (s_{1}),...,\mu^{\omega}(s_{n})  \right\}$ is the mean vector and $\Sigma^{\omega}_{ij} = \mathrm{Cov} \left\{\omega (s_{i}), \omega(s_{j})   \right\} = C^{\omega} \left\{ \omega (s_{i}), \omega(s_{j}) \right\} $ are the elements of the covariance matrix defined by the Mat\'ern  stationary isotropic covariance function. Simply, this covariance is expressed as $C^{\omega} \left(d ; \sigma^2_{\omega}, \rho_{\omega} \right)$, where $d = ||s_{i} - s_{j}||$ is the distance between two road segments; $\sigma^2_{\omega(s)}$ is the marginal variance of the spatial rutting component and $\rho_{\omega}$ is the spatial range for the spatial rutting component. These are the parameters to be estimated. 

Similarly for the annually varying spatial field $\xi_{t} (\boldsymbol{s})$, the parameters to be estimated are $\sigma^{2}_{\xi_{t}}$ and $\rho_{t}$. Detail description of the Mat\'ern  covariance structure is found in \citep[e.g.,][]{blangiardo2015spatial, krainski2018advanced} and the Supplementary Material \ref{supplementary:modelformula_SUPPLEMENTARY}. 

\vspace{-0.3cm}

\subsubsection{Other models for comparison}
\label{sec:othermodelsforcomparison}
We consider simpler non-spatial and spatial models for comparison. These models include some of the known explanatory variables and are presented in Table \ref{tab:modeltypes}. The most complex spatial model (in Equation \ref{eqn:mode4}) we refer to as {\bf Model 1}, and it's non-spatial version, Equation (\ref{eqn:model8}), as {\bf Model 4}. These models include all three explanatory variables.



\begin{tiny}
	\begin{table}[!htbp]
		\tiny
		\caption{Statistical models fitted}
		\centering
		\begin{tabular}{cccccc} 
			\hline \\ [1.5ex]
			{\bf NAME}     &  & {\bf MODEL DESCRIPTION}  &  &  &  \\ [3.5ex]
			&  & {\bf SPATIAL}   &  &  &  \\ [1.5ex]
			\cline{2-6} \\[1.5ex]
			
			{\bf Model 1}  &  &           $\eta_{t} (s) =  \beta_{\mathrm{Ac}} z_{\mathrm{Ac}(s,t) }   + \beta_{\mathrm{Agc}} z_{\mathrm{Agc}(s,t) }   + \beta_{\mathrm{Sma}} z_{\mathrm{Sma}(s,t) } + \beta_{\mathrm{d-1}} z_{\mathrm{d-1}(s,t) } +
			\beta_{\mathrm{w}} z_{\mathrm{w}(s,t) }   +  \gamma_{t} +  \xi_{t} (s)  +\omega(s) $      &  &  \\ [2.5ex]
			
			{\bf Model 2}  &  &  $\eta_{t}(s) = \beta_{\mathrm{Ac}} z_{\mathrm{Ac}(s,t) }   + \beta_{\mathrm{Agc}} z_{\mathrm{Agc}(s,t) }   + \beta_{\mathrm{Sma}} z_{\mathrm{Sma}(s,t) }   + \beta_{\mathrm{d-1}} z_{\mathrm{d-1}(s,t) }    +  \gamma_{t} + \xi_{t} (s) + \omega(s) $ &  &  \\ [2.5ex]
			
			{\bf Model 3}  &  &  $\eta_{t}(s)  = \beta_{\mathrm{Ac}} z_{\mathrm{Ac}(s,t) }  + \beta_{\mathrm{Agc}} z_{\mathrm{Agc}(s,t) } + \beta_{\mathrm{Sma}} z_{\mathrm{Sma}(s,t) }   +  \gamma_{t} + \xi_{t} (s) + \omega(s) $ &  &  \\ [5.5ex]

			&  & {\bf NON-SPATIAL}  &  &  &  \\ [1.5ex]
			\cline{2-6} \\[1.5ex]    
			
			{\bf Model 4}  &  &           $\eta_{t} (s)  = \beta_{\mathrm{Ac}} z_{\mathrm{Ac}(s,t) }   + \beta_{\mathrm{Agc}} z_{\mathrm{Agc}(s,t) }   + \beta_{\mathrm{Sma}} z_{\mathrm{Sma}(s,t) } + \beta_{\mathrm{d-1}} z_{\mathrm{d-1}(s,t) }  +
			\beta_{\mathrm{w}} z_{\mathrm{w}(s,t) } + \gamma_{t} $     &  &  \\ [2.5ex]
			
			{\bf Model 5} &  &  $\eta_{t}(s) = \beta_{\mathrm{Ac}} z_{\mathrm{Ac}(s,t) } + \beta_{\mathrm{Agc}} z_{\mathrm{Agc}(s,t) }  + \beta_{\mathrm{Sma}} z_{\mathrm{Sma}(s,t) }   + \beta_{\mathrm{d-1}} z_{\mathrm{d-1}(s,t) } + \gamma_{t} $    &  &  \\ [2.5ex]
			
			{\bf Model 6 }  &  &           $\eta_{t} (s) = \beta_{\mathrm{Ac}} z_{\mathrm{Ac}(s,t) }   + \beta_{\mathrm{Agc}} z_{\mathrm{Agc}(s,t) }  + \beta_{\mathrm{Sma}} z_{\mathrm{Sma}(s,t) }  + \gamma_{t} $    &  &  \\ [2.5ex]
			
			\hline\\[1.5ex]
		\end{tabular} 
		\label{tab:modeltypes}
	\end{table}
\end{tiny}

\vspace{-0.7cm}
\subsection{Prior models}
\label{sec:priormodels}
The priors on the explanatory variables $\boldsymbol{\beta} = (\beta_{\mathrm{Ac}}, \beta_{\mathrm{Agc}}, \beta_{\mathrm{Sma}}, \beta_{\mathrm{d-1}} ,\beta_{\mathrm{w}})$; and the hyperparameters for the random yearly effect $\gamma$ are assumed as follows
\begin{gather*}
\begin{cases}
\boldsymbol{\beta}  \stackrel{i.i.d}{\sim}\mathrm{Normal}(0, 0.001^{-1})\\    
\tau_{\gamma}\sim \mathrm{Gamma(0, \ 5.10^{-5})} \\  
\sigma_{\varepsilon} \sim \mathrm{Gamma}(0, \ 5.10^{-5}). 
\end{cases}
\end{gather*}
Vague priors are used, and we let the data inform about the parameters to be estimated. The random yearly effect is evident in density plots in Figure \ref{fig:annualrutting_density_variograms} (c), which shows the varying yearly effect of rutting. For easier interpretations, the priors for the spatial components $\xi_{t}(s)$ and $\omega (s)$ are set through the spatial range $\rho$ and marginal standard deviation $\sigma_{\omega} (s)$ of the GRFs with Mat\'ern covariance function in Equation (\ref{eqn:randge_sigma_2}). These are acquired jointly following the penalized complexity (PC) framework \citep{simpson2017penalising,fuglstad2019constructing}. The prior for the range is set through the probability $\mathrm{Pr}(\rho < 250) =0.15$, indicating that there is a 15 percent probability that the spatial range is less than 250 metres. For the marginal standard deviation, the prior  is set through the probability Pr$(\sigma_{\omega(s)} > 2.5 ) = 0.05$, indicating that there is a 5 percent probability that the marginal standard deviation for the rutting is over 2.5 millimetres. These priors are the same for all the spatial fields.

\vspace{-0.4cm}

\subsection{Inference and implementation}
\label{sec:inferenceandImplementation}
In Bayesian statistics, inference is done by computing posterior distributions for the parameters of interest based on prior models 
(Section \ref{sec:priormodels}), the likelihood model in Equation (\ref{eqn:likelihood_rutting_model_definition_1}) and the data. Our goal is to estimate the marginal posteriors of the spatial rutting component, the regression coefficients, the hyperparameters and the expected rutting (marginal likelihood), given the known explanatory variables. We use \texttt{INLA} (integrated nested Laplace approximation) \citep{rue2009approximate, rue2017bayesian,martino2019integrated}, which is implemented in the \texttt{R} package \texttt{R-INLA} to approximate these marginal distributions.

\vspace{-0.2cm}
\subsection{Model evaluation}
\label{sec:modelassessment}

Three model fit measures are available; the deviance Information Criterion (DIC) \citep{spiegelhalter2002bayesian}, the Watanabe-Akaike 
Information criterion (WAIC) \citep{watanabe2010asymptotic, gelman2014understanding} and the log-likelihood \citep{hubin2016estimating} can be obtained for each of them in order to select the best one. The DIC and WAIC are based on hierarchical modeling generalizations of the Akaike Information Criterion \citep{akaike1974new}, and  are widely used in Bayesian statistics to perform model comparison. The values are determined by the model fit and its complexity, and by a penalty applied for overfitting. Models should balance complexity and goodness-of-fit, and models with the lowest DIC and WAIC should be chosen.

\section{Results and discussion}
\label{sec:inference}
We fit the models in Section \ref{sec:othermodelsforcomparison} to the data in Section \ref{sec:variables}. For interpretability, the explanatory variables are standardized by subtracting the mean and dividing by the standard deviation. The results from the spatial Model 1 and the non-spatial version Model 4 are used to demonstrate the models' performance. Based on Model evaluation criteria (Table \ref{tab:modelcomparison_1_APPENDIX}), and for easy comparison of the models,
the spatial and non-spatial models with more explanatory variables are selected. The results for the remaining models are given in Appendix \ref{appendix:inferencefrommodelparameter_APPENDIX}. 

\subsection{Estimated spatial rutting component and yearly effect}
\label{sec:expectedannualrutting}

\begin{figure}[!htbp]
	\centering  
	\includegraphics[width=0.85\textwidth]{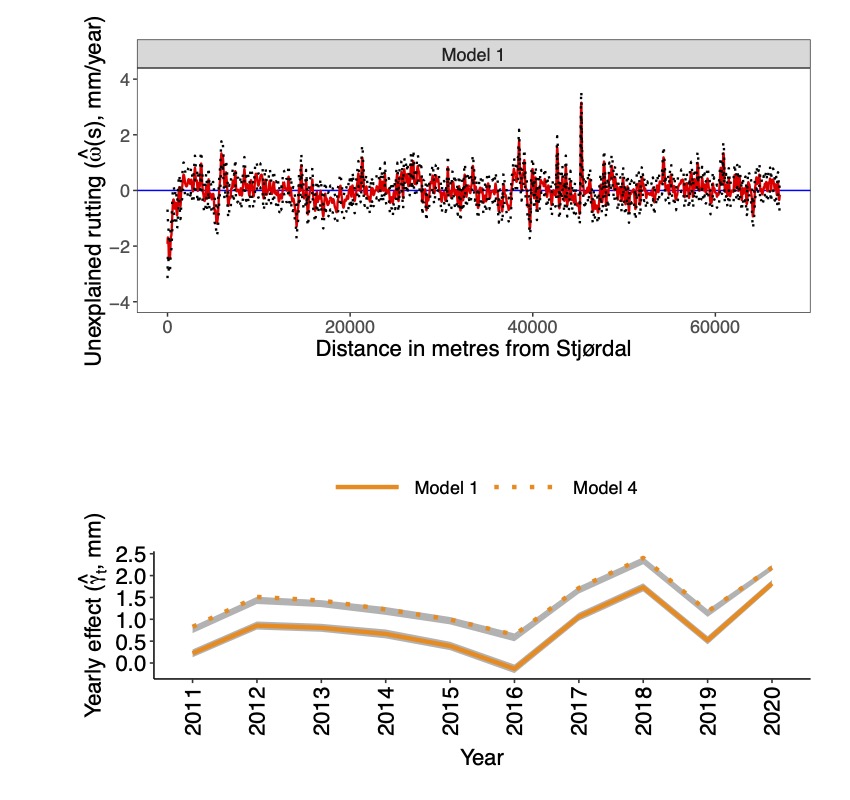} 
	
	\caption{Upper panel: Spatial rutting component (unexplained rutting,  $\hat{{\omega}} (s)$), with 95\% credible interval. Lower panel:  Rutting from the random yearly effect ($\hat{{\gamma}_{t}}$) from Model 1 (\textcolor{orange}{\rule[.5ex]{1em}{2pt}}) and Model 4 (\textcolor{orange}{\rule[.5ex]{0.1em}{3pt}} \textcolor{orange}{\rule[.5ex]{0.1em}{3pt}} \textcolor{orange}{\rule[.5ex]{0.1em}{3pt}} \textcolor{orange}{\rule[.5ex]{0.1em}{3pt}}), with 95 \% credible interval. 
		\label{fig:inferredrutting_histograms_table}}
\end{figure}

We analyze estimates from Model 1, identifying areas with high rutting rates not explained by AADT, asphalt type, and lane width. The spatial rutting component ()$\hat{\omega}(s)$) plotted in Figure \ref{fig:inferredrutting_histograms_table}(a), varies significantly along the road. Some locations exhibit up to $3.15$ mm/year extra rutting beyond what is accounted for by known factors, while others show rutting exceeding $2$ mm/year. Conversely, the minimum unexplained rutting is $-2.45$ mm/year, indicating better performance than expected.

Figure \ref{fig:inferredrutting_histograms_table}(b) presents estimates of the random yearly $\hat{\gamma}_{t}$. Rutting varies notably between years for spatial and non-spatial models, with minimum estimates in 2016 and maximum in 2018. A systematic shift between spatial and non-spatial models is evident, particularly in asphalt type coefficients (see Table \ref{tab:inference.estimates} in Section \ref{sec:estimatedmodelparameters}).

Including spatial components in the model enhances rutting variability explanation and identifies areas with unexplained high rutting rates. This is confirmed in Figure \ref{fig:inferredrutting_histograms_table} (b), where the rutting explained by the random yearly effect from the spatial model is at least $0.5$ mm/year less compared with the non-spatial model. Further discussion on estimated regression coefficients is in Section \ref{sec:estimatedmodelparameters}.

\clearpage


\subsection{Estimated model parameters}
\label{sec:estimatedmodelparameters}

The estimated parameters and 95\% credible intervals (CI) for spatial (Model 1) and non-spatial (Model 4) are shown in Table \ref{tab:inference.estimates}. Spatial models exhibit notably higher estimates for asphalt type coefficients compared with non-spatial models (see also Table \ref{fig:covariates_All_spatialnonspatial_APPENDIX} in Appendix \ref{appendix:inferencefrommodelparameter_APPENDIX}). Moreover, models lacking the spatial rutting component display larger variance in the ‘noise' term ($\hat{\sigma}{\varepsilon}$), suggesting that the spatial rutting component captures unobservable spatially varying factors such as weather and traffic. 

Additionally, the spatial model better explains expected rutting, as indicated by smaller ‘noise' terms ($\hat{\sigma}{\varepsilon}$) and uncertainty in random year effects ($\hat{\sigma}_{\gamma}$) compared with non-spatial models (refer to Table \ref{tab:inference.estimates} and Table \ref{tab:inference.estimates_APPENDIX} in Appendix \ref{appendix:inferencefrommodelparameter_APPENDIX}). Further results are provided in Supplementary Materials \ref{supplementary:addional_results}.


\clearpage

\begin{table}[!htbp]
	\scriptsize
	\caption{Estimated model parameters with $95\%$ credible intervals}
	\centering
	\setlength{\tabcolsep}{4.3pt} 
	\begin{tabular}{ccccccccccccccccccccc}
		\hline \\[0.2ex]
		& \multicolumn{2}{c}{\bf Spatial Model} & \multicolumn{2}{c}{\bf Non-spatial Model}  \\[2.2ex]
		
		& \multicolumn{2}{c}{\bf Model 1}   & \multicolumn{2}{c}{\bf Model 4} \\[1.5ex]
		\hline \\[1.0ex]
		
		& \multicolumn{1}{c}{Mean/Median} & \multicolumn{1}{c}{95\% CI} & \multicolumn{1}{c}{Mean/Median} & \multicolumn{1}{c}{95\% CI}  \\[1.0ex]
		\hline \\[1ex]  
		$\hat{\beta}_{\mathrm{Ac}}$   &  7.09  & [6.56, 7.61]  &  3.27  & [2.84, 3.71]  \\[1.2ex]
		$\hat{\beta}_{\mathrm{Agc}}$  &  7.31  & [6.08, 7.73] &  3.24  & [2.78, 3.70]    \\[1.2ex]
		$\hat{\beta}_{\mathrm{Sma}}$  &  3.53  & [3.26, 3.80] &  2.03 & [1.84, 2.22] \\[1.2ex]
		$\hat{\beta}_{\mathrm{d-1}}$ &  0.76 &  [0.68, 0.84] &  1.07  &  [0.98, 1.17]   \\[1.2ex]
		$\hat{\beta}_{\mathrm{w}}$   & 0.07 & [{0.01, 0.13}] & {0.05} & [{-0.02, 0.11}]\\[1.2ex]
		$\hat{\sigma}_{\omega (s)}$    &   0.48 &  [0.45, 0.52] & - &  - \\[1.2ex]
		$\hat{\sigma}_{\gamma}$    &   0.92 &  [0.63, 1.42]  &   1.37 &  [0.94, 2.10] \\[1.2ex]
		$\hat{\sigma}_{\varepsilon}$   &   1.38 &   [1.38, 1.39] &   1.54 &   [1.53, 1.56] \\[0.5ex]
		
		\hline \\[1ex]
	\end{tabular}
	\label{tab:inference.estimates}
\end{table}

\subsubsection{Effects of the explanatory variables}
\label{sec:fixedeffects}

Asphalt types, particularly those with weaker binder concentrations like asphalt gravel concrete ($\hat{\beta}_{\mathrm{Agc}}$) and asphalt concrete ($\hat{\beta}_{\mathrm{Ac}}$), experience more rutting (Table \ref{tab:inference.estimates}). \cite{qian2019performance} and \cite{mascarenhas2020case} shed light on how temperature and heavy traffic ($>$ 3.5 tonnes) impact rutting resistance on asphalt types such as Agc and Ac. Notably, asphalt gravel concrete is predominantly situated approximately 30 kilometers from the Swedish border, where a sub-arctic climate, underdeveloped networks, and heavy traffic converge to adversely affect pavement performance \cite{hess-11-1633-2007, ketzler2021climate, jenelius2010large, hovi2018measures, pinchasik2020crossing}. There is also greater uncertainty in asphalt concrete ($\hat{\beta}_{\mathrm{Ac}}$) and asphalt gravel concrete ($\hat{\beta}_{\mathrm{Agc}}$) compared with stone mastic asphalt ($\hat{\beta}_{\mathrm{Sma}}$), although considering spatial dependencies helps to explain more of the variability. Non-spatial models appear to explain variability between years rather than by asphalt type, suggesting a nuanced relationship between these factors in understanding rutting susceptibility.

The rut depth from the previous year ($\hat{\beta}_{\mathrm{d-1}}$) becomes increasingly important when spatial variability is not considered (Model 4, Table \ref{tab:inference.estimates}). As a genuine spatial-temporal variable, including the previous year's rut depth in the model modifies the spatial effect $\hat{\sigma}_{\omega}(s)$, specifically reducing it, which in turn enhances the precision of the estimates by decreasing uncertainty, $\hat{\sigma}_{\varepsilon}$. This improvement is evident in Models 1 and 3 (see Table \ref{tab:inference.estimates_APPENDIX} in Section \ref{appendix:inferencefrommodelparameter_APPENDIX}). Incorporating additional relevant spatial-temporal variables could further refine the results, providing a better explanation for the observed uncertainty.

The estimated effect of lane width ($\hat{\beta}_{\mathrm{w}}$) is minimal and positive, suggesting negligible impact on expected rutting.

\vspace{-0.2cm}
\subsection{Estimated rutting from the explanatory variables, random yearly effect and spatial rutting component}
\label{sec:unexplainedhotspots}

We employ Model 1 to identify maintenance ‘hot spots' where intervention is most critical. This model combines rutting inference from explanatory variable coefficients ${\hat{\beta}_{(\cdot)}}$, the mean of the yearly random effects ($\hat{\gamma}$), and the spatial rutting component ${\hat{\omega}(s)}$. Since the yearly random effects ($\gamma_{t}$) vary annually, we use an estimate of their mean, denoted as $\hat{\gamma}$. The yearly variation in ($\hat{\sigma}_{\xi_{t}}$) is small annually (Supplementary Materials \ref{supplementary:addional_results}), so it is not considered in our analysis. We account for both explained and unexplained rutting.

\begin{figure}[!htbp]
	\centering
	\subfigure[$\hat{\eta}^{*}$, $\hat{\eta}$ and mean rutting $1.62$ mm/year (\textcolor{black}{\rule[.4ex]{0.1em}{2.5pt}} \textcolor{black}{\rule[.4ex]{0.1em}{2.5pt}} \textcolor{black}{\rule[.4ex]{0.1em}{2.5pt}} \textcolor{black}{\rule[.4ex]{0.1em}{2.5pt}}).]{
		\includegraphics[width=0.43\textwidth]{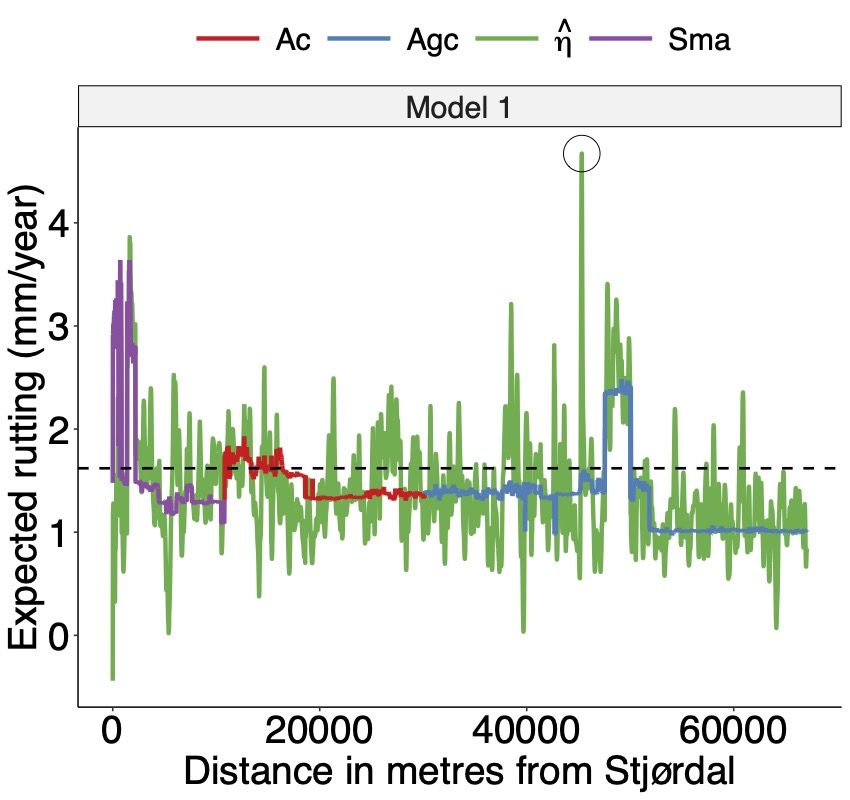}}
	\hfill
	\subfigure[$\hat{\eta}$, AADT and lane width]{
		\includegraphics[width=0.43\textwidth]{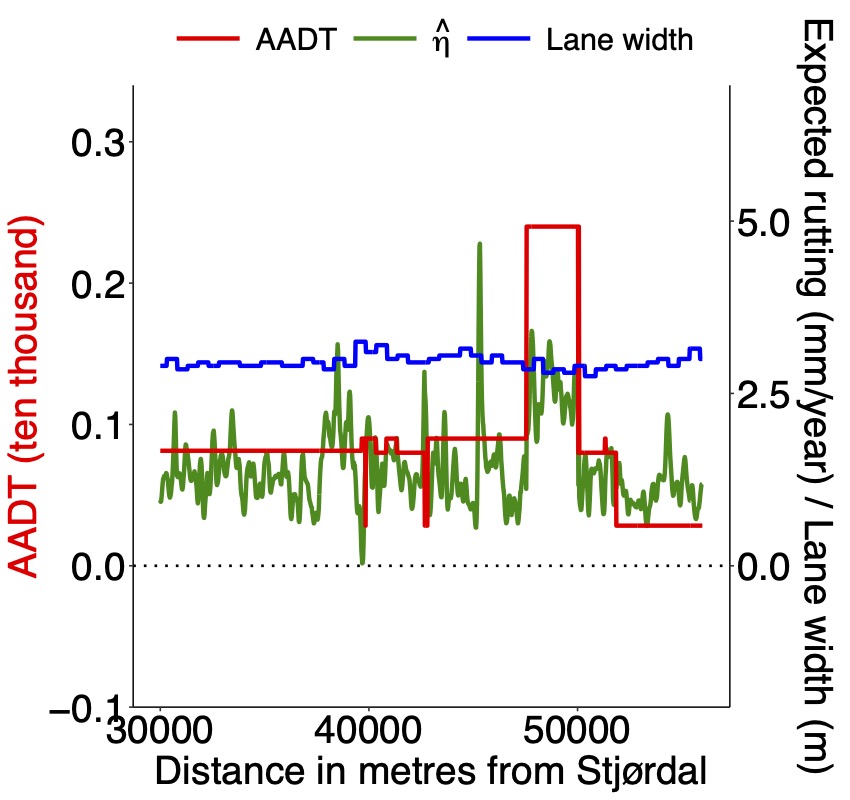}}
	
	\vspace{0.1cm}
	\begin{minipage}{\textwidth}
		\subfigure[$\hat{\eta}  > 2$ mm]{
			\includegraphics[scale=0.12]{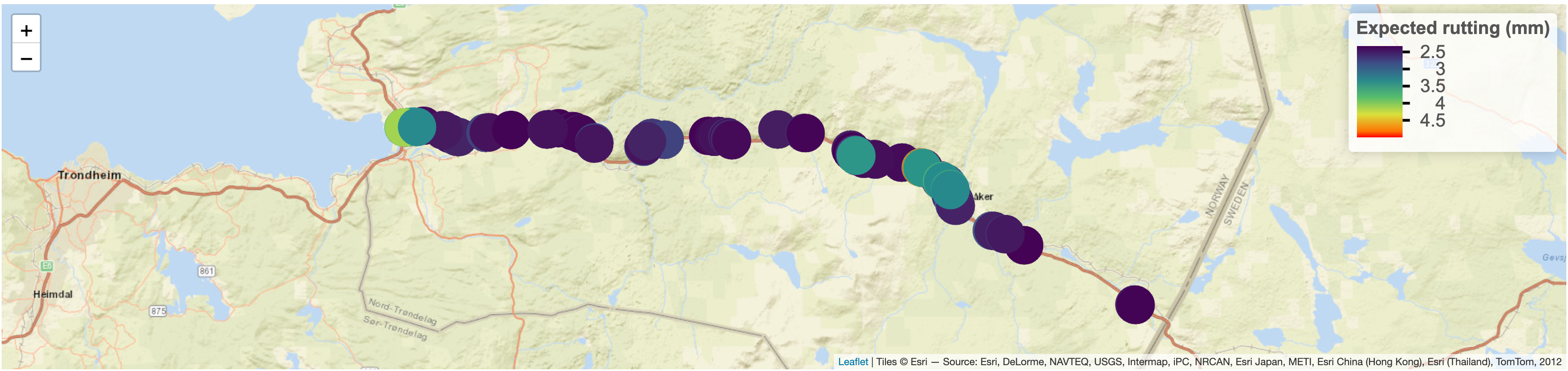}} 
		
		\hfill 
		\subfigure[$\hat{\eta}  > 3 $ mm.]{
			\includegraphics[scale=0.12]{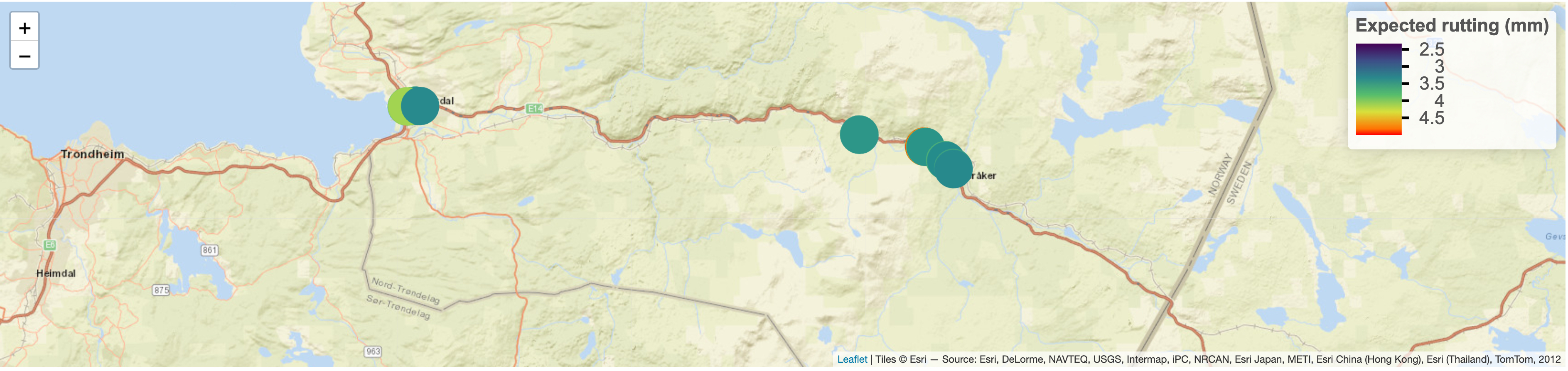}}
	\end{minipage}
	\caption{Graph (a): Expected rutting from: i)  $ \hat{\eta} $ and ii) from  $\hat{\eta}{^*} $, with each asphalt type highlighted. The effects are inferred for the year 2020. Graph (b): Expected rutting plotted against AADT and lane width for a short stretch. Graphs (c) and (d) give the locations where rutting exceeds $2$ mm/year and $3$ mm/year, respectively}. 
	\label{fig:expected_rutting_fixed_effects_spatial_effects_model1_and_4_allsections_part}
\end{figure}

Initially, rutting estimation relies on explanatory variables and the yearly random effect ($\hat{\eta}^{*} = \hat{\beta}_{{(1)}}z_{(1)(\mathrm{s,t})} + \hat{\gamma}$), with each asphalt type highlighted as Sma (\textcolor{purple}{\rule[.5ex]{1em}{2pt}}), Ac (\textcolor{red}{\rule[.5ex]{1em}{2pt}}), and Agc (\textcolor{blue}{\rule[.5ex]{1em}{2pt}}) (see Figure \ref{fig:expected_rutting_fixed_effects_spatial_effects_model1_and_4_allsections_part}, graph (a)). Incorporating spatial rutting components yields a more comprehensive estimation ${\hat{\eta} = \hat{\beta}_{(1)}z_{(1)(\mathrm{s,t})} + \hat{\gamma} + \hat{\omega}(s)}$. A comparison of these estimates highlights the influence of spatial components.

The results indicate less variability in rutting estimates without spatial components (ranging from $0.97$ to $3.64$ mm/year) compared to estimates including spatial factors (ranging from $-0.44$ to $4.67$ mm/year). This suggests that explanatory variables and the random yearly effect alone do not fully capture or represent the true rutting variability. Including spatial and yearly effects provides a more comprehensive understanding of rutting variability. Tables \ref{tab:inference.estimates} and \ref{tab:inference.estimates_APPENDIX} in Appendix \ref{appendix:inferencefrommodelparameter_APPENDIX} demonstrate this, with the remaining uncertainty ($\hat{\sigma}_{\varepsilon}$) lower in the spatial model ($1.38$ mm/year) compared to the non-spatial model ($1.54$ mm/year).

The findings indicate that high rutting estimates could arise from explanatory variables, while moderate estimates, coupled with large spatial rutting components, suggest the influence of other factors.


\clearpage


\subsubsection{Comparing expected rutting with observed data}
\label{sec:comparing_observed_with_expected}

The observed mean rutting (Figure \ref{fig:annualrutting_density_variograms} (a) in Section \ref{sec:exploratoryanalysis}) stands at $1.62$ mm/year. In Figure \ref{fig:expected_rutting_fixed_effects_spatial_effects_model1_and_4_allsections_part}, graph (a), 361 road segments have an expected rutting exceeding this mean, with 49 segments showing estimated rutting exceeding $3$ mm/year (graphs (c) and (d), respectively). These 49 locations signify potential hot spots for accelerated rutting, necessitating urgent maintenance, as their expected lifetime could be as short as 8 years, depending on traffic. Additionally, the observed mean rutting across segments ranges from -0.14 to 8.33 mm/year (Figure \ref{fig:annualrutting_density_variograms}, graph (a)), while the maximum expected rutting from the model is 4.67 mm/year (Figure \ref{fig:expected_rutting_fixed_effects_spatial_effects_model1_and_4_allsections_part}, graph (a)), approximately half the maximum observed mean rutting.

In Figure \ref{fig:expected_rutting_fixed_effects_spatial_effects_model1_and_4_allsections_part}, graph (b), we plot the expected rutting $\hat{\eta}$, lane width, and AADT (per ten thousand) for a short road section spanning road stretch 42000 to 54000 metres (or $42-54$ km from Stj\o rdal), characterized by weaker asphalt type Agc. This section exhibits spikes of accelerated rutting, particularly evident with increasing traffic. Possible factors contributing to this variability include changes in asphalt type, traffic composition, suboptimal road design, or inadequate pavement maintenance.

\subsection{Investigating an area with high rutting rates}
\label{sec:Investigating areas with unexpectedly high rutting rates}

The significant spike in Figure (\ref{fig:expected_rutting_fixed_effects_spatial_effects_model1_and_4_allsections_part}) (a), indicating high rutting rates, prompts further investigation. We examine the road stretch spanning 42000 to 54000 metres surrounding this spike, located in the municipality of Mer\r{a}ker, just 20 kilometers west of the Swedish border at Storlien. This stretch, depicted in Figure \ref{fig:expected_rutting_fixed_effects_spatial_effects_model1_and_4_allsections_part} (a), exhibits notably high rutting, peaking at $4.67$ mm, with an estimated lifespan of only $5$ years (in a low-traffic area requiring maintenance at 25 mm rut depth). Possible contributors to this accelerated rutting include clogged drainage, flooding, absence of side ditches, 
heavy traffic, or road designs featuring railguards installed close to the lane edge, which encourages traffic to remain within the same wheel tracks (refer to Figure \ref{fig:location_with_high_expected_rutting_exceeding_7mm_APPENDIX} in Appendix \ref{sec:area_with_high_rutting_APPENDIX}). Mer\r{a}ker's industrial and agricultural activities result in increased traffic from heavy vehicles, including farm equipment such as tractors (industrial, wheel drive, row crop).

\vspace{-0.4cm}

\subsubsection{Exploring the spatial model with various values of the explanatory variables}
\label{sec:model_exploration}

\begin{figure}[!htbp]
	\centering
	\subfigure[Expected rutting at road segments $42000 - 54000$ (metres) towards Sweden.]{
		\includegraphics[width=0.335\textwidth]{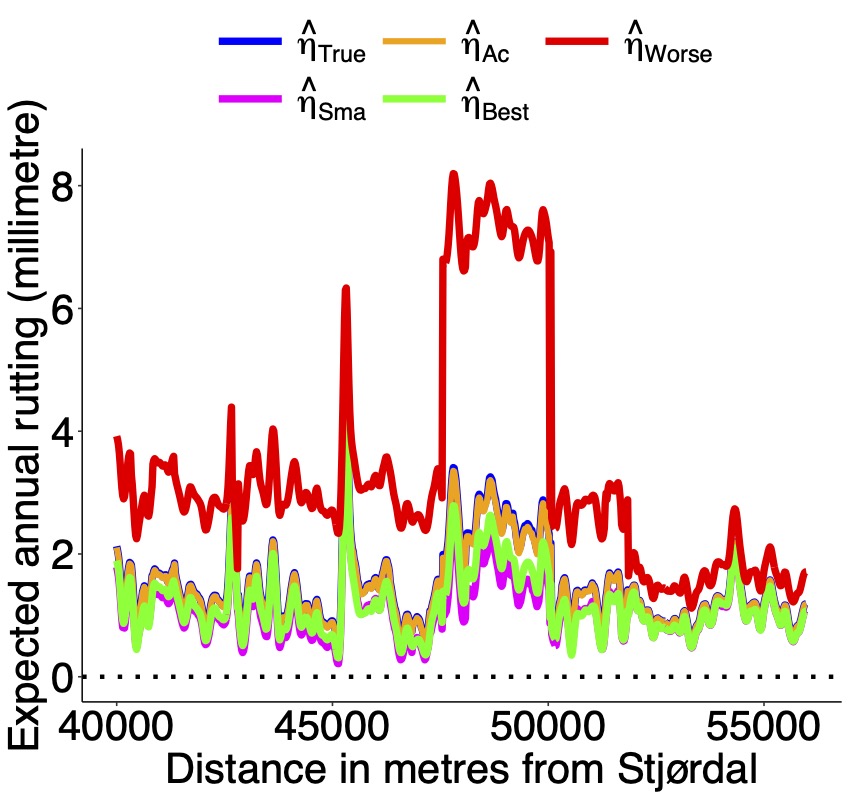}}
	\hspace{0.2cm}
	\subfigure[Expected rutting from adding each explanatory variable added sequentially.]{
		\includegraphics[width=0.335\textwidth]{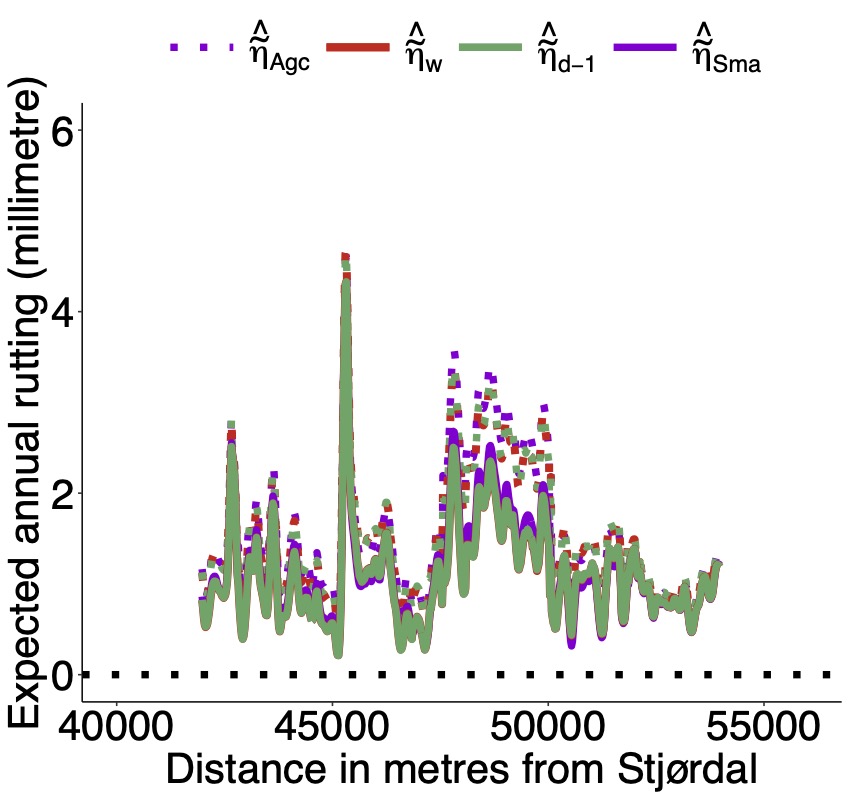}}

	\subfigure[Bearing capacity (ton).]{
		\includegraphics[width=0.305\textwidth]{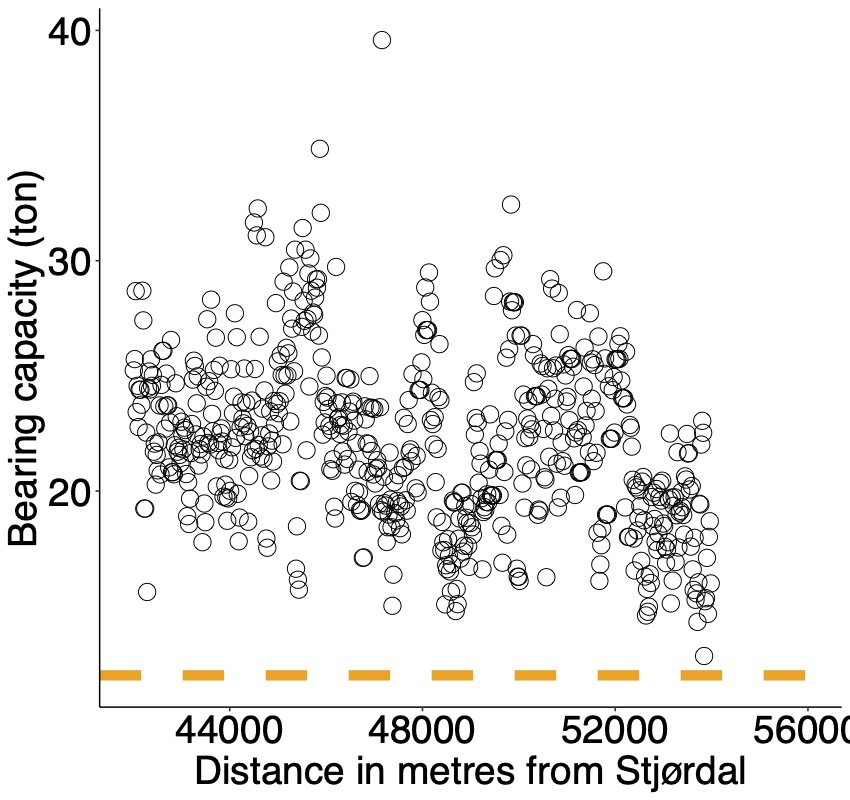}}
	\subfigure[SCI/BCI.]{
		\includegraphics[width=0.305\textwidth]{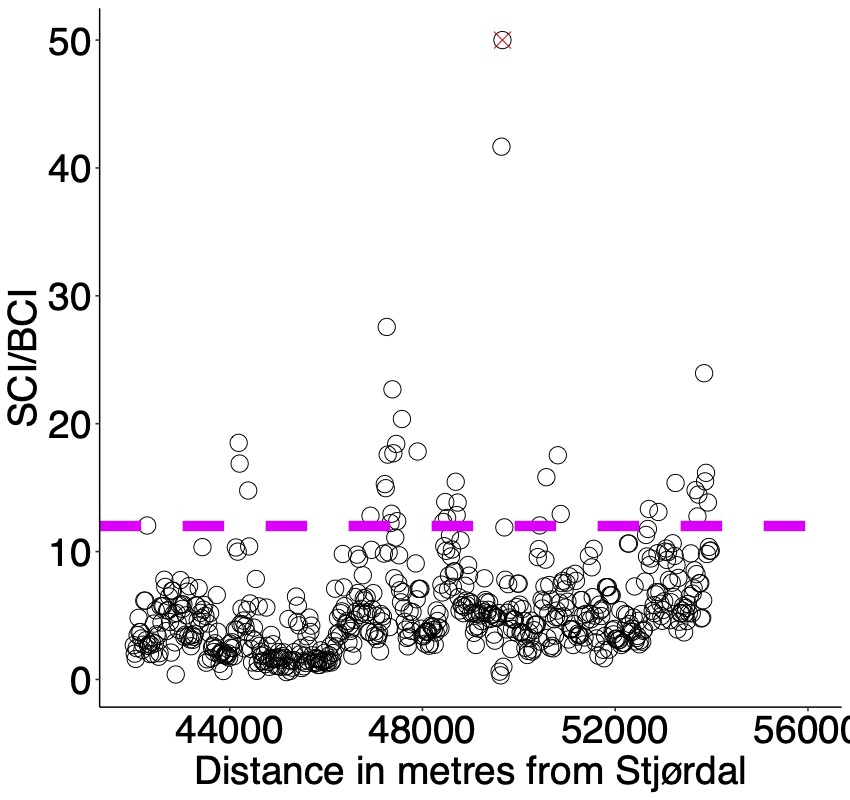}}
	\subfigure[SCI/BCI for the whole E14.]{
		\includegraphics[width=0.305\textwidth]{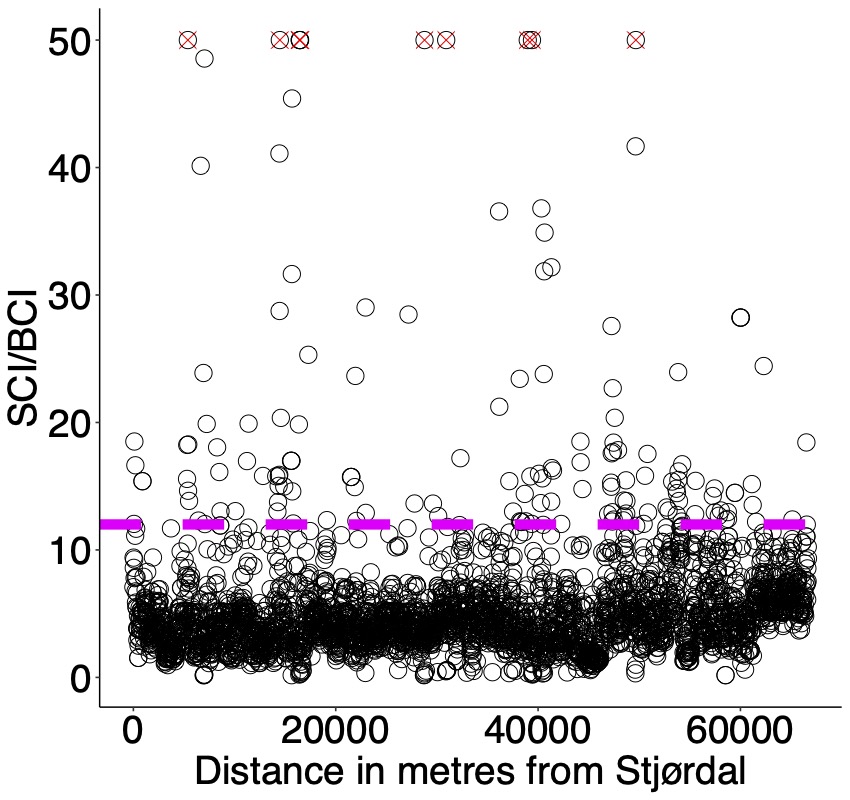}}
	
	\caption{
		Graph (a) displays expected rutting using spatial rutting component $\hat{\omega}(s)$ for road stretch 42000 to 54000 metres, along with the ‘best' or ‘worst' explanatory values. $\hat{\eta}_{\mathrm{Agc}}$ represents asphalt type Agc, random yearly effect $\hat{\gamma}$, and $\hat{\omega}(s)$. Graph (b) illustrates expected rutting inferred from $\hat{\omega}(s)$, $\hat{\gamma}$, and sequentially added explanatory variables. Graph (c) shows 2021 bearing capacity in tons; values $\ge$ 16 indicate good parameters. Graphs (d) and (e) depict SCI to BCI ratios for a short stretch or the entire E14, where values $<$ 12 indicate good curvature index. Maximum y-value plotted is 50 (values $> 50$ are denoted by the red ‘x' symbol). 
		\label{fig:expected_rutting_fixed_effects_spatial_effects_Fixedvalues_Agc_Sma}}
\end{figure}

To further investigate high rutting values in a road section, we analyze the impact of explanatory variables, yearly variation, and spatial rutting component ($\hat{\eta}$). We explore spatial Model 1 using different values for variables to estimate rutting. First, we use actual measurements like asphalt type, lane width, and rut depth ($\hat{\eta}_{\mathrm{True}}$). Second, we consider alternative asphalt types - asphalt concrete (Ac) or stone mastic asphalt (Sma) ($\hat{\eta}_{\mathrm{Ac}}$ or $\hat{\eta}_{\mathrm{Sma}}$). Third, we examine extreme values such as the more wear- and deformation-resistant pavement (Sma), minimal rut depth (0.01 mm), or wider lanes (6 meters) ($\hat{\eta}_{\mathrm{Best}}$). Conversely, accelerated rutting scenarios involve deeper rut depth (25 mm), narrower lanes (2.75 meters), and asphalt gravel concrete (Agc) ($\hat{\eta}_{\mathrm{Worse}}$) from the E14 dataset. These scenarios are depicted in Figure \ref{fig:expected_rutting_fixed_effects_spatial_effects_Fixedvalues_Agc_Sma} (a).

With worse values (\textcolor{red}{\rule[.5ex]{1em}{2pt}}), rutting is projected to be at least 1.5 mm/year ($1.76-8.20$), indicating accelerated deterioration and the need for maintenance within 3 to 14 years. Locations with unexpectedly high rutting, exceeding 4 mm/year, are observed, possibly due to increased AADT compared to other areas (Figure \ref{fig:expected_rutting_fixed_effects_spatial_effects_model1_and_4_allsections_part}, graph b).

Using the best covariate values (\textcolor{green}{\rule[.5ex]{1em}{2pt}}), rutting is projected to be at most 4.30 mm/year, with maintenance required approximately 6 years after construction or maintenance. Additionally, using the asphalt binder Sma yields similar rutting estimates but generally lower than those using the best values, indicating asphalt type as a primary factor. Asphalt gravel concrete (Agc, \textcolor{blue}{\rule[.5ex]{1em}{2pt}}) and asphalt concrete (Ac, \textcolor{yellow}{\rule[.5ex]{1em}{2pt}}) show similar rutting estimates due to similar binder concentrations.


\clearpage

We examine the impact of asphalt type on rutting (Figure \ref{fig:expected_rutting_fixed_effects_spatial_effects_Fixedvalues_Agc_Sma}, graph (b)). Changing from asphalt type Agc (asphalt gravel concrete) to the more wear- and deformation-resistant Sma (stone mastic asphalt) while keeping all other variables constant, we sequentially add additional explanatory variables to the model, starting with asphalt type. Initially, asphalt type (Agc or Sma) is included with the random yearly effect $\hat{\gamma}$ and the spatial rutting component $\hat{\omega}(s)$ as follows: $\hat{\tilde{\eta}}_{\mathrm{Agc}} = \hat{\beta}_{\mathrm{Agc}}z_{\mathrm{Agc (s,t)}} + \hat{\gamma} + \hat{\omega}(s)$ or $\hat{\tilde{\eta}}_{\mathrm{Sma}} = \hat{\beta}_{\mathrm{Sma}}z_{\mathrm{Agc (s,t)}} + \hat{\gamma} + \hat{\omega}(s)$. Subsequently, the previous year's rut depth is incorporated, represented as $\hat{\tilde{\eta}}_{\mathrm{d-1}} = \hat{\beta}_{\mathrm{Agc}}z_{\mathrm{Agc (s,t)}} + \hat{\beta}_{\mathrm{d-1}}z_{\mathrm{d-1 (s,t)}} + \hat{\gamma} + \hat{\omega}(s)$ or $\hat{\tilde{\eta}}_{\mathrm{d-1}} = \hat{\beta}_{\mathrm{Sma}}z_{\mathrm{Sma (s,t)}} + \hat{\beta}_{\mathrm{d-1}}z_{\mathrm{d-1 (s,t)}} + \hat{\gamma} + \hat{\omega}(s)$. Similarly, the model with added lane width ${\hat{\beta}_{w}z_{\mathrm{w (s,t)}} }$ is denoted as $\hat{\tilde{\eta}}_{\mathrm{w}}$. Results from using stone mastic asphalt (Sma) are depicted by solid lines, and asphalt gravel concrete (Agc) results are shown with dotted lines. 

The findings affirm that asphalt type significantly influences rutting. Implementing the binder type Sma instead of the actual asphalt type Agc would decrease expected rutting, reaching a maximum of $4.35$ mm/year (compared to 4.68 mm/year), approximately reducing rutting by 0.33 mm/year annually. Over a projected lifetime of 6 years, this equates to an additional year of serviceability.

\subsubsection{Evaluating pavement subgrade support on expected rutting}
\label{sec:evaluating_subgrade_support}

We examine bearing capacity, surface curvature index (SCI, micrometre), and base curvature index (BCI) for road segments 42000-54000 using 2021 data, indicative of construction quality and soil pressure sustainability over the past decade. Bearing capacity is a measure of the deflection caused by a rolling load axle on pavement. It is measured using the Rapid Pavement Tester (Raptor, \url{https://www.ramboll.com/rst/rst-products}), which continuously collects data on the pavement structure. A value  exceeding 16 metric tons (\textcolor{orange}{\rule[.5ex]{0.1em}{3pt}} \textcolor{orange}{\rule[.5ex]{0.1em}{3pt}} \textcolor{orange}{\rule[.5ex]{0.1em}{3pt}} \textcolor{orange}{\rule[.5ex]{0.1em}{3pt}}) suggests a robust foundation design (Figure \ref{fig:expected_rutting_fixed_effects_spatial_effects_Fixedvalues_Agc_Sma}, graph (c)). The SCI to BCI ratio  (Figure \ref{fig:expected_rutting_fixed_effects_spatial_effects_Fixedvalues_Agc_Sma}, graphs (d) and (e)) reveals deflection location within the pavement layers. High rutting may result from factors like side slope height (data unavailable), inadequate drainage, or sub-base layer deformation during thawing periods. Bearing capacity data, sampled and temperature-corrected during summer, may inflate measurements compared to thawing periods. Varying construction material quality along the road stretch (30000-67100 metres) warrants investigation into surface and subsoil layer deflections. An SCI/BCI ratio below 12 (\textcolor{magenta}{\rule[.5ex]{0.1em}{2pt}} \textcolor{magenta}{\rule[.5ex]{0.1em}{2pt}} \textcolor{magenta}{\rule[.5ex]{0.1em}{2pt}} \textcolor{magenta}{\rule[.5ex]{0.1em}{2pt}}) indicates robust construction. Inconsistent ratios suggest layer deformation, with data indicating uneven construction on E14, featuring deeper base layer deflections.

In conclusion, asphalt type significantly influences rutting, recommending more wear resistant and deformation resistant binder types for extended lifespan.

\section{{How road authorities can make use of this model}}
\label{sec:discussion}

The spatial-statistical model serves as a potent tool for road authorities, offering a data-driven approach to enhance maintenance strategies, identify critical areas for intervention, and optimize road infrastructure durability. This approach facilitates a more precise diagnosis of rutting, enabling authorities to make informed decisions for sustainable road infrastructure.

Our analyses reveal that rutting exceeding 2 mm/year is expected at several locations along E14. Given the expected lifetime of the pavement and recommended treatment levels, this level of rutting is concerning. Maintenance carried out at rut depths of 20-25 mm, depending on traffic, and pavement designs expected to last between 20-30 years indicate that some sections of the road may only survive for at most 12.5 years (10 years in high-traffic areas). It's imperative to consider spatial rutting components; models fitted with variables known to influence rutting and random yearly effects alone underestimate rutting variability, thus inflating pavement lifetime estimates. Including the spatial rutting component reduces estimated pavement lifetimes to at most 6 years, compared to 5 years without it. This information is crucial for road authorities' long-term planning and resource allocation.

We've demonstrated the spatial-statistical model's ability to identify areas with unexpectedly high rutting, pinpointing hot spots where rutting is accelerating and indicating urgent maintenance needs. By analyzing factors influencing rutting, random effects, and spatial components, authorities can prioritize resources for targeted interventions. Additionally, the model provides a comprehensive understanding of rutting, enhancing authorities' ability to make informed decisions about maintenance priorities and pavement materials during construction or rehabilitation.

The model can investigate contributing factors to rutting beyond asphalt type, such as drainage issues, heavy traffic, and road design flaws. Comparisons with non-spatial models emphasize the importance of spatial factors in explaining rutting variations, reinforcing the need for spatial models for accurate inference and predictions. The estimated model parameters and credible intervals provide a robust foundation for data-driven decision-making, enabling road authorities to make informed choices regarding maintenance priorities and infrastructure improvements.

\section{Conclusion}
\label{sec:conclusion}
This research provides valuable insights into the rutting phenomenon by elucidating both well-established factors and those that remain unexplained. It demonstrates the importance of incorporating spatial information to accurately estimate rutting variability, showing that observations in close proximity exhibit higher similarity than those at a distance. To achieve a comprehensive understanding, it is imperative to consider both observed quantities as well as unobserved quantities. Due to Norway's specific climatic context, where temperatures rarely exceed 30 $^\circ \mathrm{C}$ and annual averages remain below 20 $^\circ \mathrm{C}$, temperature was not included in our rutting model. The maximum average mean temperature per month is less than 18 $^\circ \mathrm{C}$, indicating that temperature extremes, which could significantly impact rutting by accelerating asphalt softening, are infrequent. Therefore, we focused on more relevant factors such as asphalt type, prior rut depth, and traffic load, which exhibit significant spatial and temporal variability. This approach provides a more accurate and contextually appropriate understanding of the rutting phenomenon in Norway.

The study confirms the significant impact of asphalt type selection on rutting performance, with rut-resistant mixtures proving more effective in mitigating rutting. It also highlights the influence of lane width on pavement rutting. Narrower roads experience more concentrated traffic loads, leading to increased rutting compared to a wider road that is otherwise similar (i.e., same AADT, asphalt type and rut depth levels). However, wider roads, while distributing traffic loads over a larger area, tend to experience higher rut depths, likely due to increased traffic volumes and higher wheel loads.

To address rutting and minimize its effects over time, integrating a life-cycle cost-benefit perspective into road design and planning processes is advisable. Further research should explore additional factors influencing rutting, such as bearing capacity, ditch depth for water outflow, and lane width, to enhance the credibility of research results and model applicability to other road networks. Additionally, evaluating the long-term performance of rut-resistant asphalt mixtures under various traffic and environmental conditions would be beneficial. This comprehensive approach ensures that our model provides more meaningful insights and practical recommendations for mitigating rutting in similar climatic regions.

	\section*{Acknowledgement}
The authors are grateful to J\o rn Vatn and Jochen K\"ohler at the Norwegian University of Science and Technology for invaluable discussions on analysis and interpretation.

\section*{Disclosure Statement}
The authors report no declarations of interest.

\section*{Funding}
This work was supported by the Norwegian University of Science and Technology in collaboration with Statens Vegvesen Project [grant number 9056608 – Smarter Maintenance (2019-2023)].

\bibliographystyle{tfv}
\bibliography{interacttfvsample}

\vspace{-0.4cm}

\section{Appendices}

\appendix

\renewcommand\thesection{\Alph{section}}
\renewcommand{\thefigure}{\thesection.\arabic{figure}}
\setcounter{figure}{1}
\renewcommand{\thefigure}{\thesection.\arabic{figure}}

\setcounter{figure}{1}
\renewcommand{\thefigure}{\thesection.\arabic{figure}}

\section{Inference from model parameters}
\label{appendix:inferencefrommodelparameter_APPENDIX}
Table \ref{tab:inference.estimates_APPENDIX} compares estimates from the proposed spatial and non-spatial models. The spatial models account for spatially varying uncertainties, such as weather and explanatory variables, e.g., bearing capacity. Including more explanatory variables reduces the noise term ($\hat{\sigma}_{\varepsilon}$) and the random yearly effect ($\hat{\sigma}_{\gamma}$).


\begin{table}[!htbp]
	\tiny
	\caption{Estimated model parameters with $95\%$ credible intervals}
	\label{tab:inference.estimates_APPENDIX}
	\begin{adjustbox}{width=\textwidth,totalheight=\textheight,keepaspectratio,rotate=0,
			nofloat=table}
		\setlength{\tabcolsep}{2.3pt} 
		\begin{tabular}{ccccccccc}
			\hline \\[1ex]
			\multicolumn{7}{c}{\bf SPATIAL MODELS} & \\[1.5ex]
			
			& \multicolumn{2}{c}{\bf Model 1} &
			\multicolumn{2}{c}{\bf Model 2} & \multicolumn{2}{c}{\bf Model 3}  \\[2.5ex]
			\hline \\[1ex]
			
			& \multicolumn{1}{c}{Mean/Median} & \multicolumn{1}{c}{95\% CI} & \multicolumn{1}{c}{Mean/Median} & \multicolumn{1}{c}{95\% CI}& \multicolumn{1}{c}{Mean/Median} & \multicolumn{1}{c}{95\% CI}  \\[1ex]
			\hline \\[1ex]  
			$\hat{\beta}_{\mathrm{Ac}}$  &  7.09 & [6.56, 7.61] & 7.08 &[6.55, 7.61] & 6.89 & [6.34, 7.44]   \\[1ex]
			$\hat{\beta}_{\mathrm{Agc}}$ &  7.31 & [6.88, 7.73]&  7.29 & [6.86, 7.71] & 7.07 & [6.64, 7.50]   \\[1ex]
			$\hat{\beta}_{\mathrm{Sma}}$ &  3.53 &[3.26, 3.80] & 3.55 & [3.28 3.82] & 3.31 & [3.03, 3.59]   \\[1ex]
			$\hat{\beta}_{\mathrm{d-1}}$ &  0.76 &[0.68, 0.84] & 0.76  &[0.68, 0.84] & - & -   \\[1ex]
			$\hat{\beta}_{\mathrm{w}}$   & 0.07 &[0.01, 0.13] & - &-  & -  \\[1ex]
			$\hat{\sigma}_{\omega (s)}$    &  0.48 & [0.45, 0.52]& 0.48 &[0.45, 0.52] & 0.52 & [0.48, 0.57]   \\[1ex]
			$\hat{\sigma}_{\gamma}$    & 0.92 & [0.63, 1.42] & 0.93 & [0.63, 1.40]& 0.93 & [0.63, 1.42]  \\[1ex]
			$\hat{\sigma}_{\varepsilon}$  & 1.38 &[1.38, 1.39] & 1.38 & [1.38, 1.39]  & 1.39 &[1.38, 1.40]    \\[3.5ex]
			
			\multicolumn{7}{c}{\bf NON-SPATIAL MODELS} & \\[2.2ex]
			
			&  \multicolumn{2}{c}{\bf Model 4} & 
			\multicolumn{2}{c}{\bf Model 5} & \multicolumn{2}{c}{\bf Model 6}  \\[2.5ex]
			\hline \\[1.5ex]
			
			& \multicolumn{1}{c}{Mean/Median} & \multicolumn{1}{c}{95\% CI} & \multicolumn{1}{c}{Mean/Median} & \multicolumn{1}{c}{95\% CI}& \multicolumn{1}{c}{Mean/Median} & \multicolumn{1}{c}{95\% CI}  \\[1.2ex]
			\hline \\[1ex]
			$\hat{\beta}_{\mathrm{Ac}}$  &  3.27 & [2.84, 3.71] & 3.28 & [2.85, 3.72]& 2.72 & [2.29, 3.16]      \\[1ex]
			$\hat{\beta}_{\mathrm{Agc}}$ &  3.24 &[2.78, 3.70] &  3.23 &  [2.77, 3.70] & 2.71 & [2.25, 3.17]  \\[1ex]
			$\hat{\beta}_{\mathrm{Sma}}$ &  2.03 &[1.84, 2.22] & 2.05  &[1.86, 2.24] & 1.55 & [1.36, 1.73]   \\[1ex]
			$\hat{\beta}_{\mathrm{d-1}}$ &  1.07 & [0.98, 1.17] & 1.08 & [0.98, 1.18] &- & -  \\[1ex]
			$\hat{\beta}_{\mathrm{w}}$   & {0.05} & [{ -0.02, 0.11}] & - & - & - & -  \\[1ex]
			$\hat{\sigma}_{\gamma}$    & 1.37 & [0.94, 2.10] &  1.37 & [0.94, 2.10] & 1.41 & [0.96, 2.17]   \\[1ex]
			$\hat{\sigma}_{\varepsilon}$  & 1.54 & [1.53, 1.56] & 1.56 &[1.54, 1.57] & 1.56 & [1.54, 1.57]   \\[1ex]
			\hline \\[1ex]
		\end{tabular}
	\end{adjustbox}
\end{table}


\clearpage
\subsection{Model Comparison}
\label{appendix:model_comparison_APPENDIX}

Model comparison criteria (Table \ref{tab:modelcomparison_1_APPENDIX}) select the models with more explanatory variables most often as the best model, with the highest goodness-of-fit (lowest DIC and WAIC). As shown in the results above (e.g., Table \ref{tab:inference.estimates_APPENDIX}), the more complex spatial models (Model 1) and non-spatial models (Model 4 or 5) capture the random and spatial variability of pavement rutting and explains more of the unstructured error terms with the addition of explanatory variable lane width.


\begin{center}
	\begin{table}[!htbp]
		\scriptsize
		\caption{Model comparison}
		\setlength{\tabcolsep}{2.5pt}
		\begin{tabular}{cccccccccccccccccc}
			\hline \\[1.5ex]
			{\bf Model}     & \multicolumn{3}{c}{\bf Estimates from spatial models} &   {\bf Model} & \multicolumn{3}{c}{\bf Estimates from non-spatial models}     \\[1ex]
			\cline{2-4} \cline{6-9}\\[1ex]
			& DIC       & WAIC     & (-) loglikelihood  && DIC       & WAIC     & (-) loglikelihood   \\[1.5ex]
			\cline{2-4} \cline{6-9}\\[1.5ex]
			
			{ \bf 1 } & {\bf 195067.76} &  { \bf 195660.91} & { -101464.72  }&  {\bf 4}    &  { \bf 104290.61} &  {\bf 100996.40}  &  { -52229.53}        &   \\[1ex] 
			
			{\bf 2} & 195154.12 &  {195669.17} &  {\bf -101460.32 }  &  {\bf 5}  & { 104290.62} & { 100996.45} & {\bf -52223.62}      &          \\[1ex]
			
			{\bf 3} & {195242.58} &  {195959.89}  & -101624.69  & {\bf 6}   &  104763.10   & 101556.00 & -52453.66       &       \\[1ex]

			\hline \\[0.2ex]
		\end{tabular}
		\label{tab:modelcomparison_1_APPENDIX}
	\end{table}
\end{center}


\vspace{-1.6cm}

\subsection{Marginal Posterior Distributions of Explanatory Variables}
\label{appendix:marginalposterior_explanatoryvariables_APPENDIX}

\begin{center}
	\begin{figure}[!htbp]
		\centering
		\subfigure[Asphalt type]{ 
			\includegraphics[width=0.415\linewidth]{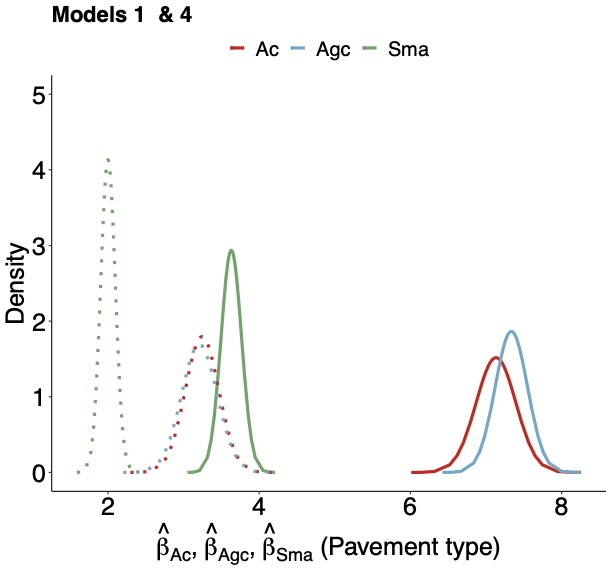}} 
		\label{fig:model4pavement.sec}
		\subfigure[Rut depth]{ 
			\includegraphics[width=0.415\linewidth]{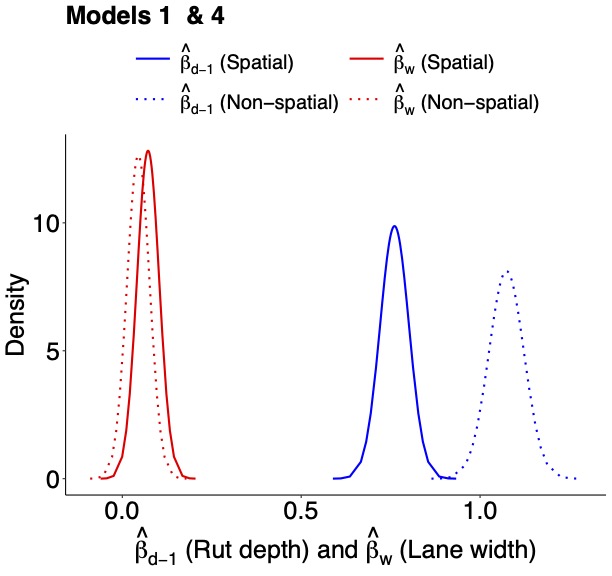}}
		\label{fig:model4rut_depth.sec}
		\hfill
		\caption{Marginal posterior distributions of asphalt type (solid lines are the spatial models, and the dotted lines are the non-spatial models), rut depth from the previous year and lane width.}
		\label{fig:covariates_All_spatialnonspatial_APPENDIX}
	\end{figure}
\end{center}

Figure \ref{fig:covariates_All_spatialnonspatial_APPENDIX} gives the marginal posterior distributions of asphalt type, rut depth from the previous year and lane width. Narrow bands of the distributions indicate that the uncertainty is smaller. Graph (a) shows a clear shift in asphalt type's contribution to rutting from the spatial and non-spatial models. This is because asphalt type is spatially varying. Stone mastic asphalt (Sma) gave lower rutting values because it has a more wear resistant binder. 

\vspace{-1.2cm}
\subsection{Marginal Posterior Distributions of Random Effects}
\label{appendix:marginalposterior_randomeffects_APPENDIX}

\begin{center}
	\begin{figure}[!htbp]
		\begin{minipage}{\textwidth}
			\centering
			\subfigure[Marginal posterior distributions of the standard deviation of the Gaussian observations $\hat{\sigma}_{\varepsilon}$.]{
				\includegraphics[width=0.32\textwidth]{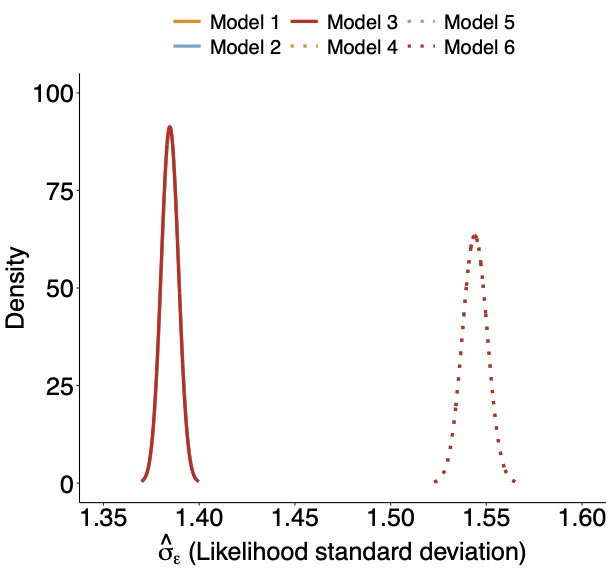}}
			\hspace{0.35cm}
			\subfigure[Marginal posterior distributions of the standard deviation of the random yearly effect $\hat{\sigma}_{\gamma}$.]{
				\includegraphics[width=0.32\textwidth]{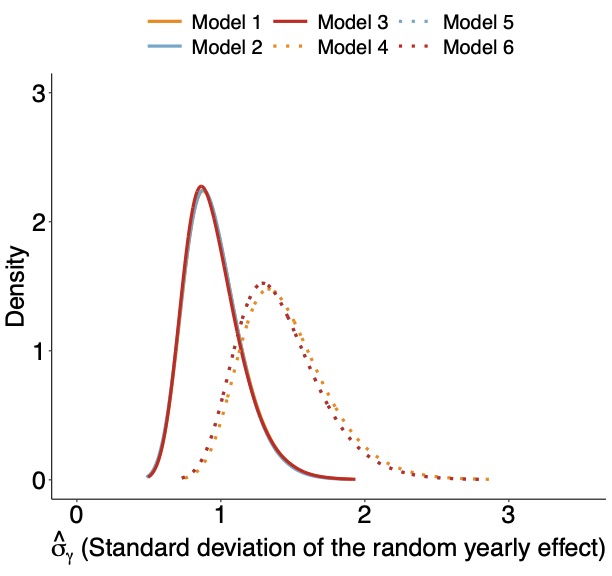}}

			\vspace{0.35cm}
			\subfigure[Marginal posterior distributions of the nominal range of the common spatial field $\hat{\rho}$ .]{
				\includegraphics[width=0.32\textwidth]{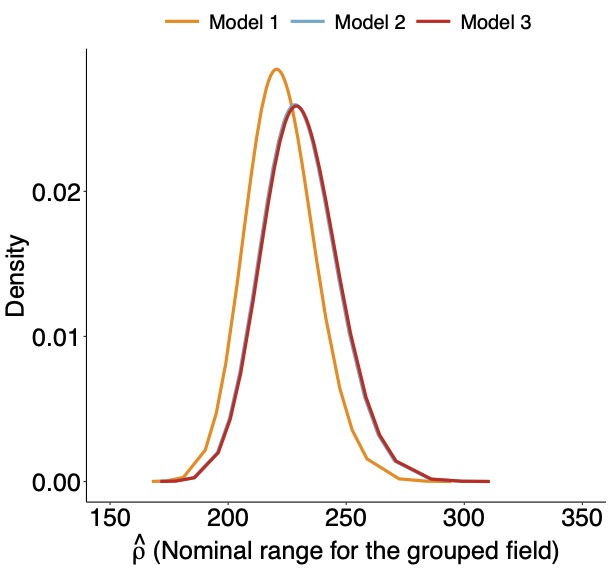}}
			\hspace{0.35cm}
			\subfigure[Marginal posterior distributions of the standard deviation of the common spatial field  $\hat{\sigma}_{\omega(s)}$.]{
				\includegraphics[width=0.32\textwidth]{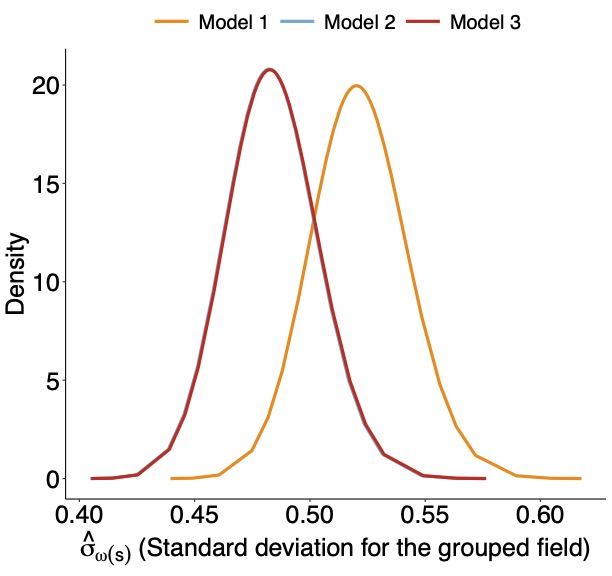}}
		\end{minipage}
		\caption{Estimates of marginal posterior distributions of the spatial models (solid lines) and non-spatial models (dotted lines).}
		\label{fig:marginal_posterior_Gaussian_year_range_standard.alternative_APPENDIX}
	\end{figure}
\end{center}

Figure \ref{fig:marginal_posterior_Gaussian_year_range_standard.alternative_APPENDIX} presents the marginal posterior distributions of the standard deviation of Gaussian observations ($\hat{\sigma}_{\varepsilon}$), random yearly effect ($\hat{\sigma}_{\gamma}$), range of the common spatial field ($\hat{\rho}$), and standard deviation of the common spatial field ($\hat{\sigma}_{\omega(s)}$). Solid and dotted lines represent spatial and non-spatial models, respectively, with narrower densities indicating smaller estimated uncertainties. The spatial model captures some of the uncertainties in Gaussian observations and yearly effects, suggesting that unobserved variability (e.g., weather) is partly explained by the spatial rutting component. This is evidenced by smaller noise term estimates ($\hat{\sigma}_{\varepsilon}$) in the spatial model (Table \ref{tab:inference.estimates_APPENDIX}).

As more explanatory variables are added (e.g., rut depth), the variability of the spatial field ($\hat{\sigma}_{\omega(s)}$) decreases (Figure \ref{fig:marginal_posterior_Gaussian_year_range_standard.alternative_APPENDIX}, graph (d)). In Model 2, including previous year's rut depth gives a large effect, indicating deeper ruts lead to increased rutting (Figure \ref{fig:covariates_All_spatialnonspatial_APPENDIX}, graph (b)). The spatial range ($\hat{\rho}$) is generally larger in Model 2 compared to Model 3, likely due to varying weather patterns, but including lane width reduces the effect of rut depth on expected rutting.

\subsection{Area with high rutting rates}
\label{sec:area_with_high_rutting_APPENDIX}

\begin{figure}[!htbp]
	\centering
	\subfigure[Short section between \texttt{EV14 S6D1 m6802-m6882 F1} with high rutting.]{
		\includegraphics[width=0.365\textwidth]{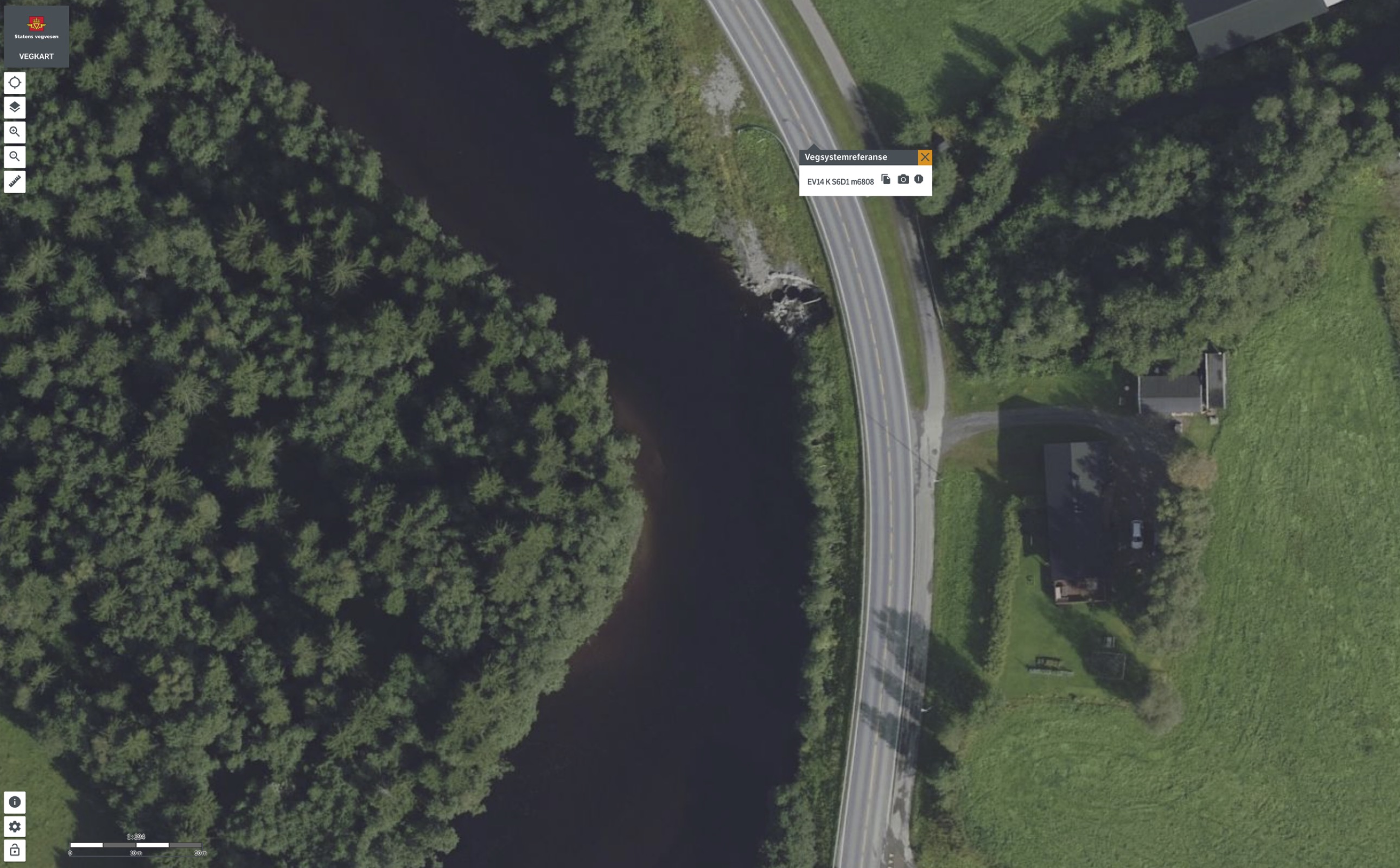}}
	\hspace{0.2cm}
	\subfigure[Location \texttt{EV14 S6D1 m6881 F1} in year 2020]{
		\includegraphics[width=0.365\textwidth]{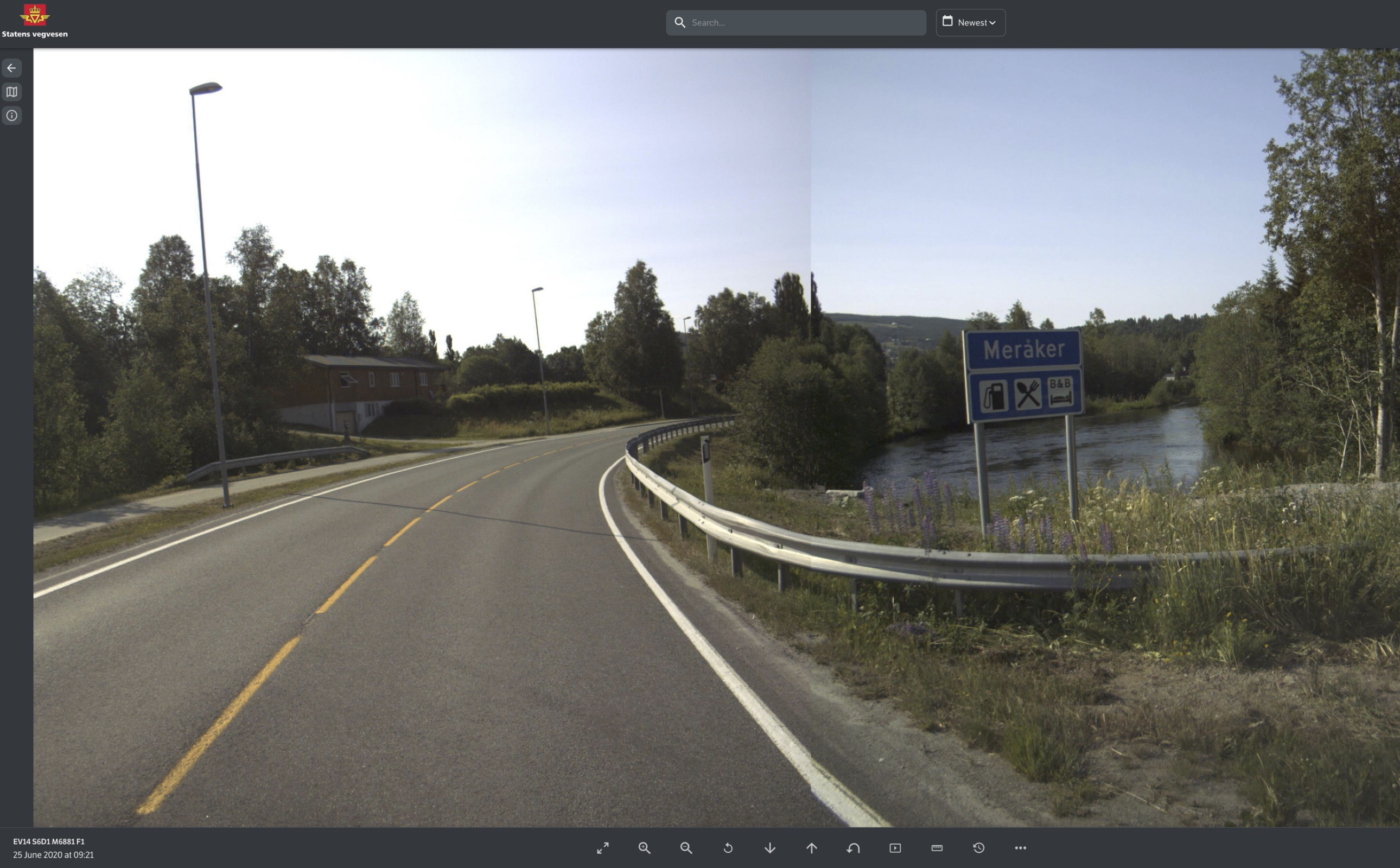}}
	\vspace{0.5cm}
	\subfigure[Location \texttt{EV14 S6D1 m6808 F1} in year 2021]{
		\includegraphics[width=0.365\textwidth]{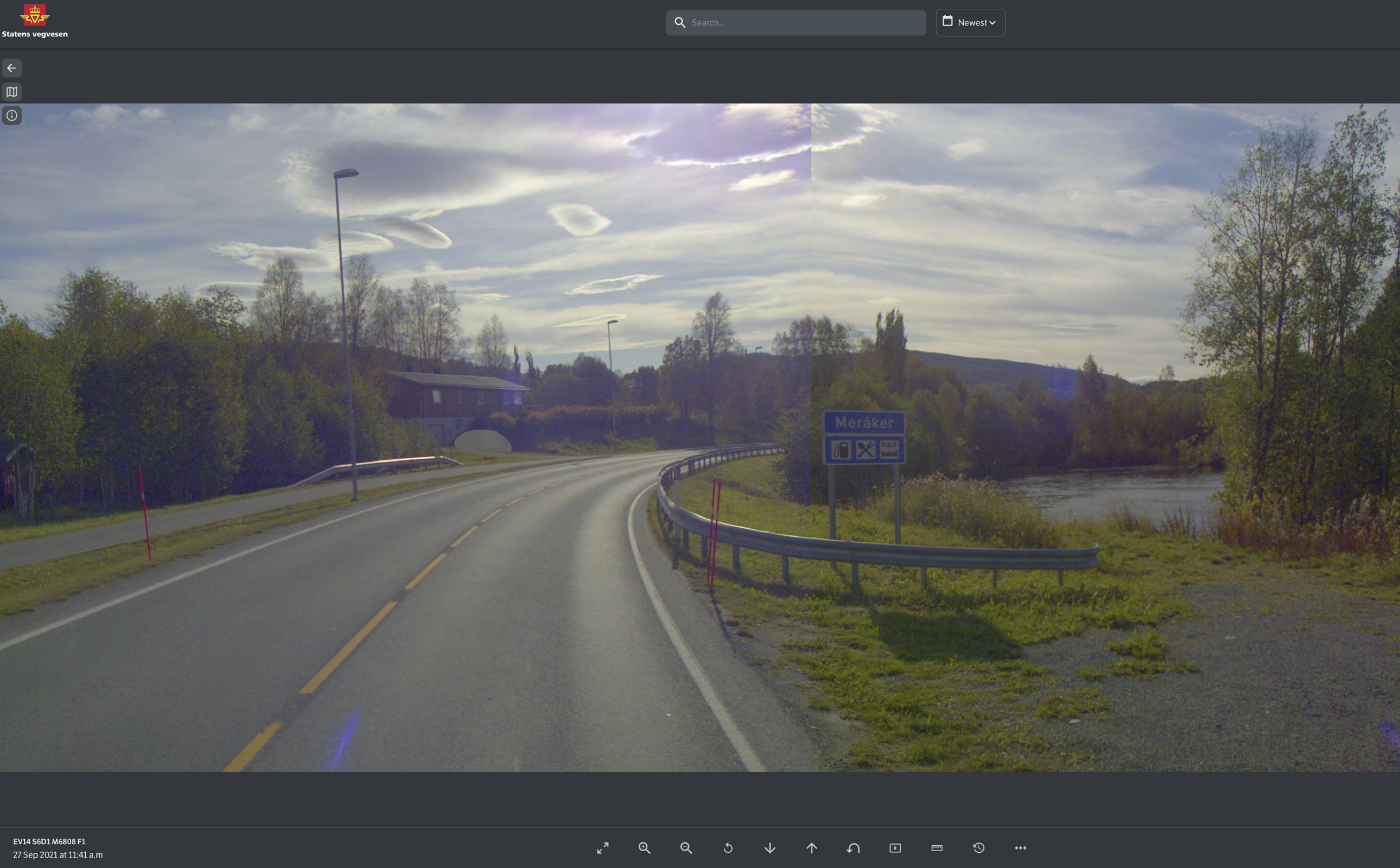}}
	\hspace{0.2cm}
	\subfigure[Location \texttt{EV14 S6D1 m6807 F1} in year 2022]{ 
		\includegraphics[width=0.365\textwidth]{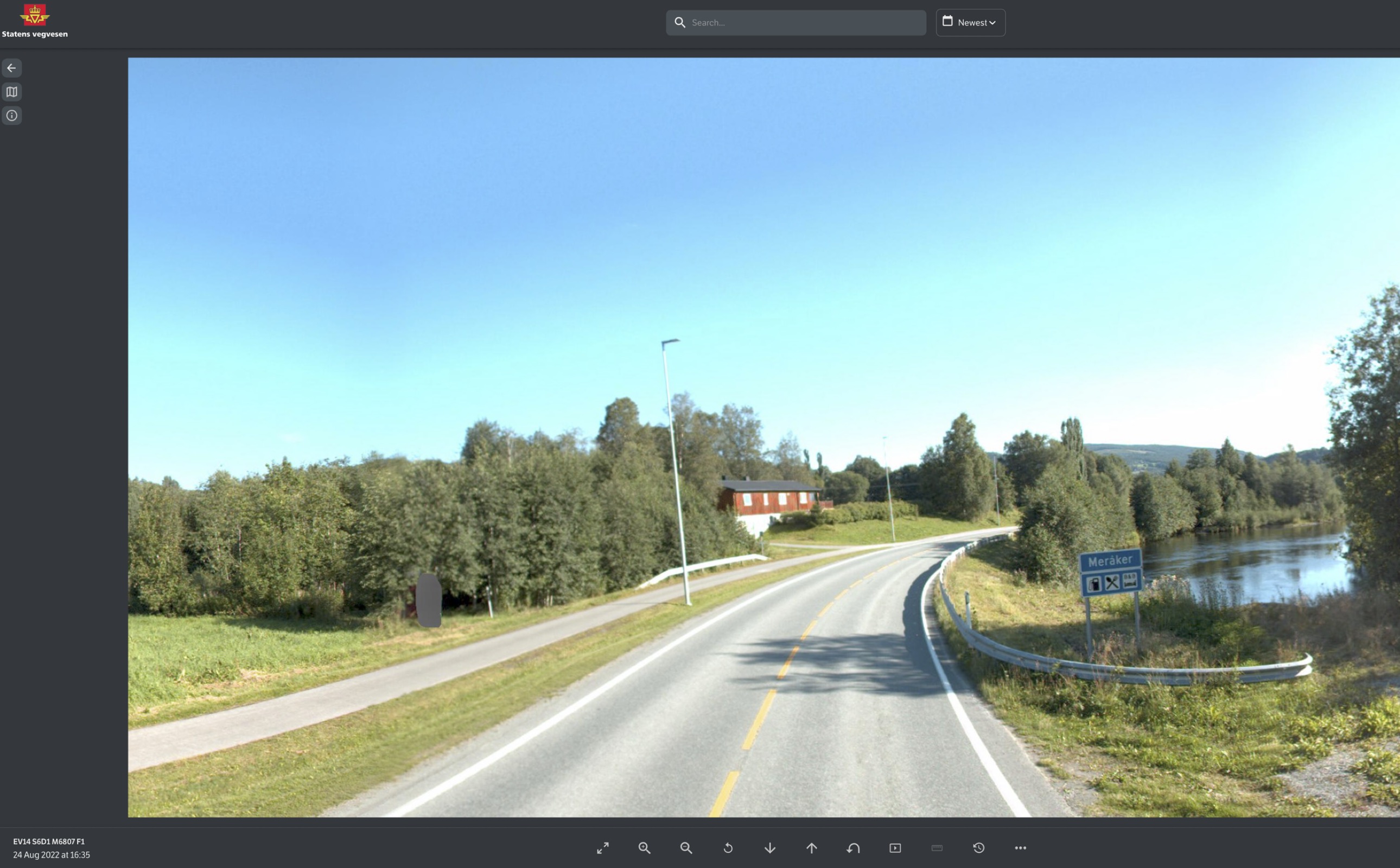}}
	\caption{Graph (a): Location with high rutting rates (road segments $2100-2700$ towards Sweden). Graphs (b), (c) and (d) show the location, in years 2020-2022 (see 
		\href{https://vegbilder.atlas.vegvesen.no/?lat=63.42546005&lng=11.73406457&view=image&zoom=16&imageId=Vegbilder_2020.2020-06-25T09.21.58.907073_EV00014_S6D1_m06881&year=2020}{Statens vegvesen} or go to the next url: \url{https://vegkart.atlas.vegvesen.no/\#kartlag:nib/@337039,7037105,18/vegsystemreferanse:337039.29:7037149.25}}
	\label{fig:location_with_high_expected_rutting_exceeding_7mm_APPENDIX}
\end{figure}

Figure \ref{fig:location_with_high_expected_rutting_exceeding_7mm_APPENDIX} shows the location where rutting exceeds $7$ mm when the worse values of the explanatory variables are used to explore Model 1 (see Section \ref{sec:Investigating areas with unexpectedly high rutting rates})


\subsection{Average mean air surface temperature}
\label{sec:temperature_appendix}

In Norway, the temperature rarely exceeds 30 $^\circ \mathrm{C}$, and on average, monthly temperatures do not exceed 20 $^\circ \mathrm{C}$. This limited variation in high temperatures suggests that temperature may not play a critical role in the rutting process within this specific climatic context. The maximum average mean temperature per month or over years in Norway is less than 18 $^\circ \mathrm{C}$ (and less than $13.5 \  ^\circ \mathrm{C}$ for all years, except 2104 and 2018). This indicates that temperature extremes, which could significantly influence rutting, are infrequent.

\clearpage

\begin{figure}[!htbp]
	\centering
	\includegraphics[width=0.50\textwidth]{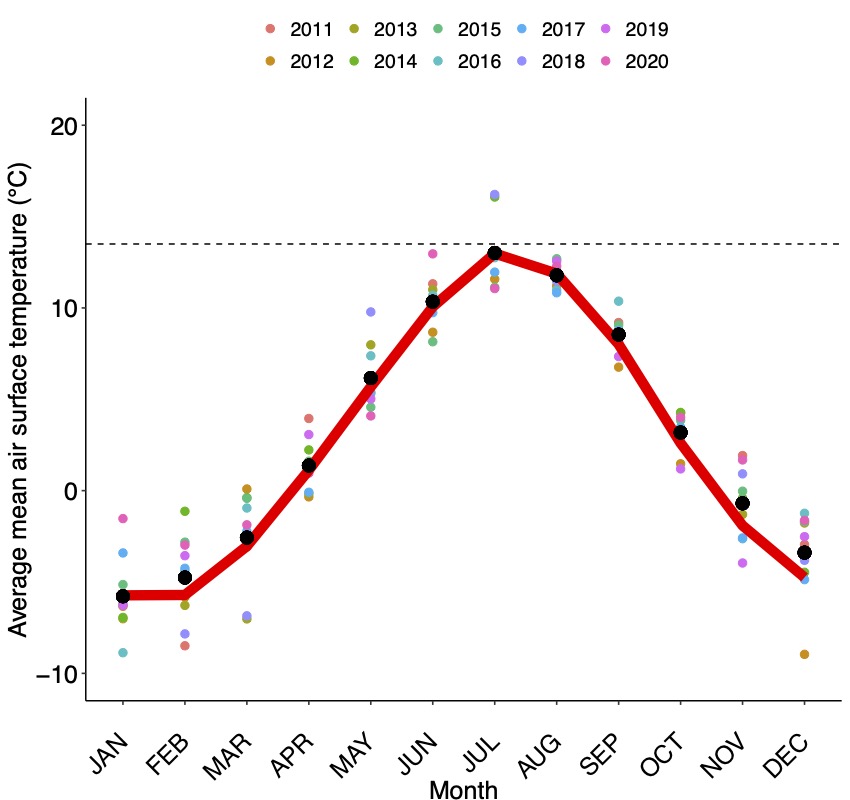}
	\caption{Average mean air surface temperature ($^\circ \mathrm{C}$,  \textcolor{red}{\rule[.5ex]{1em}{2pt}}) for years 2011-2020. The black circles ($\bullet$) gives the average mean monthly air surface temperature and the coloured dots indicated in the legend gives the mean monthly for each year. The dashed black line gives a temperature value of $13.5 ^\circ \mathrm{C}$ (see \url{ https://climateknowledgeportal.worldbank.org/country/norway/trends-variability-historical} for further details).}
	\label{fig:temperature_APPENDIX}
\end{figure}

\clearpage

\end{document}


	\label{Natoya}
	\title{Supplementary Material: A Spatial-statistical model to analyse historical rutting data.}
	
\author{
	\name{N.~O. A. S Jourdain\textsuperscript{a}\thanks{CONTACT N.~O. A. S. Jourdain. Email: natoya.jourdain@ntnu.no}, I. Steinsland\textsuperscript{a}, M. Birkhez-Shami\textsuperscript{a}, E. Vedvik\textsuperscript{c},   W. Olsen\textsuperscript{d}, D. Gryteselv\textsuperscript{b}, D. Siebert\textsuperscript{b} and A. Klein-Paste\textsuperscript{a}}
	\affil{\textsuperscript{a}Norwegian University of Science and Technology (NTNU), Trondheim, Norway; \\ \textsuperscript{b}Statens Vegvesen, Trondheim, Norway; \\
		\textsuperscript{c}Fremtind Forsikring, Trondheim, Norway;
	}
}
	
	\maketitle

\section{Data summary}
\label{supplementary:datasummary}
\begin{figure}[!htbp]
	\begin{minipage}[!ht]{0.6\linewidth} 
            \tiny
		\setlength{\tabcolsep}{4pt}
         \subfigure[Summary of rut data]{
		\begin{tabular}[1ht]{p{0.9cm} p{0.3cm}p{0.7cm}p{0.7cm}p{0.3cm}p{0.3cm}p{0.3cm}p{0.3cm}p{0.3cm}p{0.3cm}p{0.3cm}}
			\hline \\[1.2ex] 
			Year & \multicolumn{1}{c}{Missing rut} & \multicolumn{1}{c}{Rut}  & \multicolumn{1}{c}{Width}   & \multicolumn{1}{c}{Slope} & \multicolumn{1}{c}{Min. rutting}  & \multicolumn{1}{c}{Max. rutting}    \\[1.5ex] 
			\hline \\ [1ex] 
			2010   & 167    & 3188 & 3338 & 816 & 0 & 0  \\[1.3ex] 
			2011   & 1      & 3354 & 3338 & 816 & $-11.1$ & 40.9  \\[1.3ex]  
			2012   & 7      & 3348 & 3338 & 816 & $-8.9$ & 202.2\\[1.3ex] 
			2013   & 0      & 3355 & 3338 & 816 & $-10.7$ & 36.7 \\[1.3ex] 
			2014   & 1089   & 2257 & 3338 & 816 & $-9.3$ & 19.6  \\[1.3ex] 
			2015   & 17     & 3338 & 3338 & 816 & $-8.9$ & 30.5  \\[1.3ex] 
			2016   & 0      & 3355 & 3338 & 816 & $-9.6$ & 19.7  \\[1.3ex] 
			2017   & 0      & 3355 & 3338 & 816 & $-5.4$ & 17.2  \\[1.3ex] 
			2018   & 0      & 3355 & 3338 & 816 & $-4.5$ & 20.4 \\[1.3ex] 
			2019   & 0      & 3355 & 3338 & 816 & $-6.0$ & 15.4    \\[1.3ex] 
			2020   & 0      & 3355 & 3338 & 816 & $-5.9$ & 10.1     \\[1.3ex] 
			\hline \\[1ex] 
		    \end{tabular}}
            \end{minipage}\hfill
	\begin{minipage}{0.4\linewidth}
		\centering
          \subfigure[AADT data]{
		\includegraphics[width=0.8\textwidth]{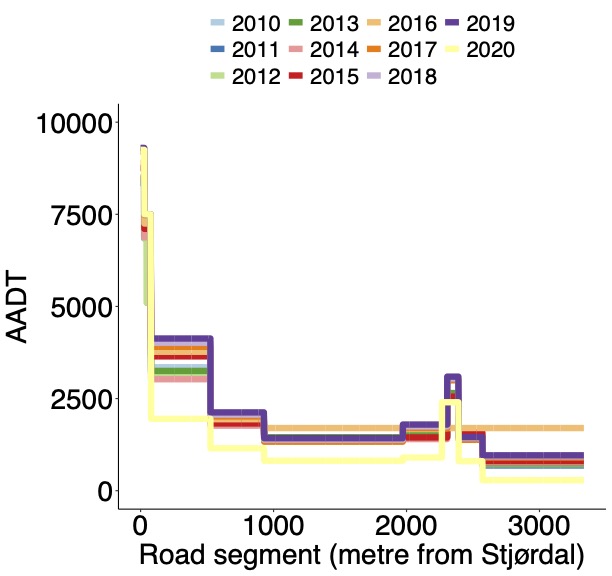}}
	\end{minipage}
         \caption{Available rut depth, road width, slope height measurements, and minimum and maximum rutting are shown. Year 2014 has ($\sim$) 32\% NaNs for rut depth, suggesting that re-pavement has taken place at these locations. This means that the correlation between rut depth values will be fast deteriorating with distance, suggesting a small spatial range process.}
		\label{fig:mapofnorway_data_1_APPENDIX}
\end{figure}

Figure \ref{fig:mapofnorway_data_1_APPENDIX} gives a summary of the available data used in the analysis. There are a large number of missing values in year 2014. This suggests that repavement took place in this year. The AADT is given in graph (b). The year 2020 gave, in general, the lowest AADT values. This is a result of the COVID-19 pandemic.

\vspace{-0.5cm}

\section{Mat\'ern covariance structure}
\label{supplementary:modelformula_SUPPLEMENTARY}

The annually varying spatial process $\xi_{t} (\boldsymbol{s})$ is a Gaussian random process  $\left\{\xi_{t} (\boldsymbol{s}) : \boldsymbol{s} \in \mathcal{D} \subseteq \mathbb{R}^{d} \right\} $ such that for any $n\ge 1$ and for each set of spatial locations $\left(s_{1},...,s_{n}  \right)$ satisfies 
\begin{center}
	\begin{equation}
	\boldsymbol{\xi}_{t} (\boldsymbol{s}) = \left\{{\xi}_{t} (s_{1}),..., {\xi}_{t} (s_{n}) \right\} = \left({\xi}_{t,1},..., {\xi}_{t,n}     \right) \sim \mathrm{Normal} \left(\boldsymbol{\mu}, \Sigma \right).
	\label{eqn:annuallyvaryingspatialfield}
	\end{equation}
\end{center}
Here $\boldsymbol{\mu} = \left\{\mu (s_{1}),...,\mu(s_{n})  \right\}$ is the mean vector and $\Sigma_{ij} = \mathrm{Cov} \left\{\xi_{t} (s_{i}), \xi_{t}(s_{j})   \right\} = C \left\{ \xi_{t} (s_{i}), \xi_{t}(s_{j}) \right\} $ are the elements of the covariance matrix defined by the Mat\'ern  stationary isotropic covariance function
\begin{center}
	\begin{equation}
	C \left\{ \xi_{t} (s_{i}), \xi_{t}(s_{j}) \right\} = \frac{\sigma^{2}_{\xi_{t}}}{2^{\nu-1} \Gamma(\nu)}\left(\kappa ||s_{i} -s_{j}||_{2} \right)^{\nu} K_{\nu} \left(\kappa ||s_{i} -s_{j}||_{2} \right),
	\label{eqn:covarianceannuallyfield}
	\end{equation}
\end{center}
for $s_{i}, s_{j} \in \mathcal{D}$, where $|| \cdot||_{2}$ denotes the Euclidean distance. The scaling parameter is $\kappa >0$; $K_{\nu}$ is the modified Bessel function of second order kind and order $\nu = \alpha - d/2>0$ such that $\alpha = \nu + d/2$, which controls the smoothness of the realisations; and $\sigma^2_{\xi_{t}}$ is the marginal variance defined as: 

$$\sigma^2_{\xi_{t}} = \frac{\Gamma(\nu)} {(4 \pi)^{d/2} \Gamma(\alpha) \kappa^{2 \nu}\tau^2}. $$

Similarly, the common spatial process $\omega(\boldsymbol{s})$ is a Gaussian random process $\left\{\omega (\boldsymbol{s}) : \boldsymbol{s} \in \mathcal{D} \subseteq \mathbb{R}^{d} \right\} $ defined as 
\begin{center}
	\begin{equation}
	\omega(\boldsymbol{s}) = \left\{{\omega} (s_{1}),..., {\omega} (s_{n}) \right\} = \left({\omega}_{1},..., {\omega}_{n}     \right) \sim \mathrm{Normal} \left(\boldsymbol{\mu}^{\omega}, \Sigma^{\omega} \right),
	\label{eqn:commonspatialfield}
	\end{equation}
\end{center}
{where $\boldsymbol{\mu}^{\omega} = \left\{\mu^{\omega} (s_{1}),...,\mu^{\omega}(s_{n})  \right\}$ is the mean vector and $\Sigma^{\omega}_{ij} = \mathrm{Cov} \left\{\omega (s_{i}), \omega(s_{j})   \right\} = C^{\omega} \left\{ \omega (s_{i}), \omega(s_{j}) \right\} $ are the elements of Mat\'ern  stationary isotropic covariance function defined in Equation (\ref{eqn:covarianceannuallyfield}), but with marginal variance defined as $\sigma^2_{\omega(s)}$; the scaling parameter given as $\kappa_{\omega} >0$; and with $K_{\nu_{\omega}}$ defined as the modified Bessel function of second order kind and order $\nu_{\omega}= \alpha_{\omega} - d/2 >0$.} 

Following \cite{lindgren2011explicit}, $\omega(\boldsymbol{s})$ and $\xi_{t}(\boldsymbol{s})$  are modelled with a stationary SPDE (stochastic partial differential equation) approach \citep{ingebrigtsen2014spatial, blangiardo2015spatial, ingebrigtsen2015estimation, krainski2018advanced}, with form 
\begin{center}
	\begin{equation}
	\mathcal{W}(\boldsymbol{s})= (\kappa^2 -\Delta)^{\alpha/2} (\tau \xi_{t}(\boldsymbol{s})).
	\label{eqn:spdeannual}
	\end{equation}
\end{center}
or
\begin{center}
	\begin{equation}
	\mathcal{W}_{\omega}(\boldsymbol{s})= (\kappa^2_{\omega} -\Delta)^{\alpha_{\omega}/2} (\tau_{\omega} \omega(\boldsymbol{s})).
	\label{eqn:spdecommon}
	\end{equation}
\end{center}
{respectively, where 
	$\Delta$ is the Laplacian; $\tau$ (or $\tau_{\omega}$) controls the variance; and $\mathcal{W}(\boldsymbol{s})$ or $\left\{ \mathcal{W}_{\omega}(\boldsymbol{s}) \right\}$ is the Gaussian spatial white noise}. The solutions to the SPDEs are the stationary Gaussian random fields $\xi_{t}(\boldsymbol{s})$ and  $\omega(\boldsymbol{s})$. With this SPDE approach, the Mat\'ern field is represented by a linear combination of basis functions defined in a triangulation of a given spatial domain $\mathcal{D}$ \citep{juan2021enhancing}.  
In this paper, the spatial domain along Highway road E14  is assumed to be one-dimensional ($d=1$), with a triangulation having vertices every $\mathrm{20^{th}}$ metre or one road segment. 

 Instead of using the scaling parameter $\kappa_{\omega}$ (or $\kappa$),  it is common to use the spatial range $\rho$ for the common spatial field (or $\rho_{t}$ for the annually varying spatial field) defined empirically as $\rho = 2 \sqrt{2 \nu_{\omega}/\kappa_{\omega}}$. Here, we assume the default value  $\alpha_{\omega}=2$ in \texttt{R-INLA}, such that $\nu_{\omega} =3/2$, and the commonly used spatial range $\rho$ and marginal variance of the spatial field $\sigma^2_{\omega(s)}$: 
\begin{center}
	\begin{equation}
	\rho = \frac{2\sqrt{3}}{\kappa_{\omega}}
	\quad\text{and}\quad 
	\sigma^2_{\omega(s)} = \frac{1}{4 \kappa^3_{\omega}\tau^2_{\omega}}.
	\label{eqn:randge_sigma_2}
	\end{equation}
\end{center}

\section{Additional Results}
\label{supplementary:addional_results}
Here, we give some additional results of the inference and random effects obtained from the spatial models.

\subsection{Inference from the explanatory variables}
\label{appendix:inferencefromfixedeffects}

\subsubsection{Pavement type}
\label{sec:pavementtype_appendix}

\begin{figure}[!h]
 \subfigure[]{
	\includegraphics[width=0.32\textwidth]{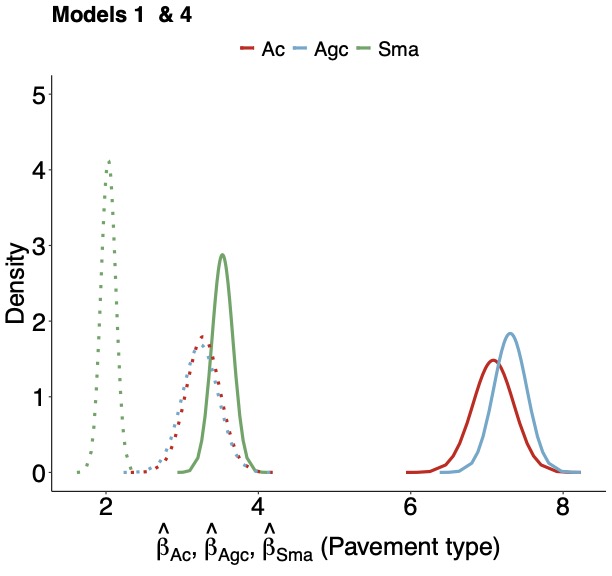}}
	\label{fig:model2pavement}
 \subfigure[]{
	\includegraphics[width=0.32\textwidth]{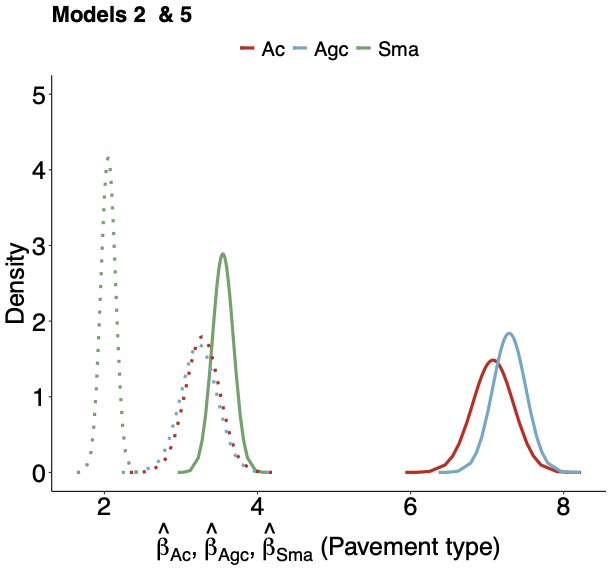}}
	\label{fig:model3pavement}
	%
 \subfigure[]{
	\includegraphics[width=0.32\textwidth]{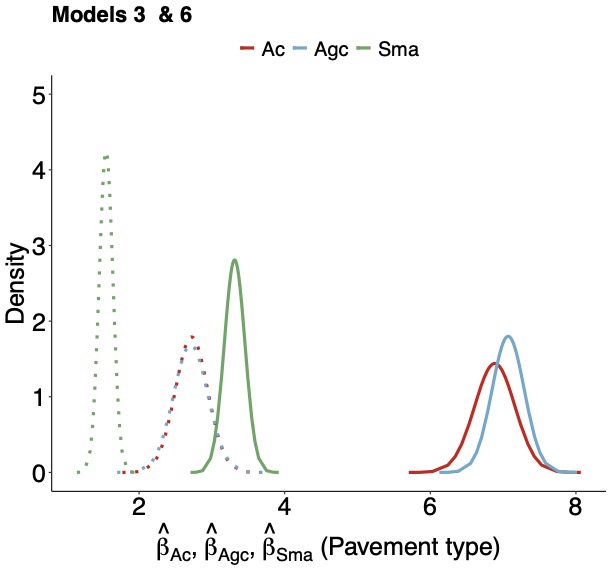}}
	\label{fig:model4pavement1}
	\caption{Marginal posterior distributions of pavement type. Solid lines are the spatial models and the dotted lines are the non-spatial models.}
	\label{fig:covariates_pavement_type_spatialnonspatial_appendix}
\end{figure}

The fixed effect of pavement type on expected annual rutting from all models is given in Figure \ref{fig:covariates_pavement_type_spatialnonspatial_appendix}. The effect of asphalt concrete (Ac) and asphalt gravel concrete (Agc) on expected annual rutting is very clear when the spatial effects are considered (solid lines). Also, the uncertainty from using either of the pavement types is also clearly identified, with stone mastic asphalt type (Sma) having the smallest variability. When the spatial effects of pavement type is not considered, the effects on rutting are not easily distinguishable (dotted lines). 


\subsubsection{Rut depth and Lane width}
\label{sec:rutdepth_appendix}
\begin{figure}[!h]
    \subfigure[]{
	\includegraphics[width=0.305\textwidth]{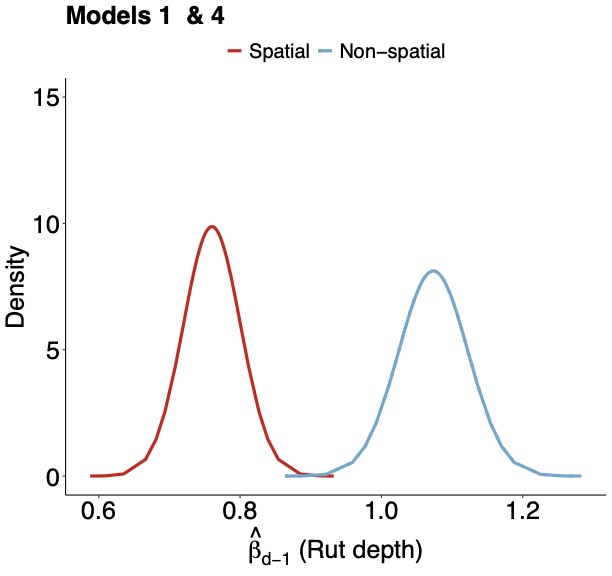}}
	\label{fig:model1rut_depth}
	\hfill
	 \subfigure[]{
	\includegraphics[width=0.305\textwidth]{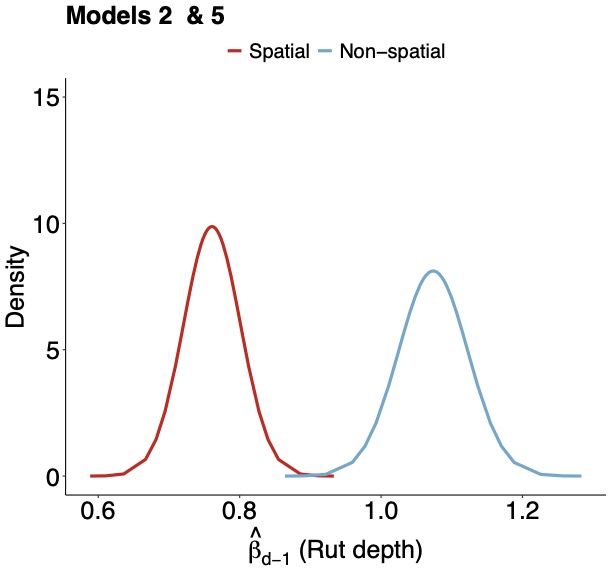}}
\label{fig:model2rut_depth}
    \hfill
 \subfigure[]{
	\includegraphics[width=0.305\textwidth]{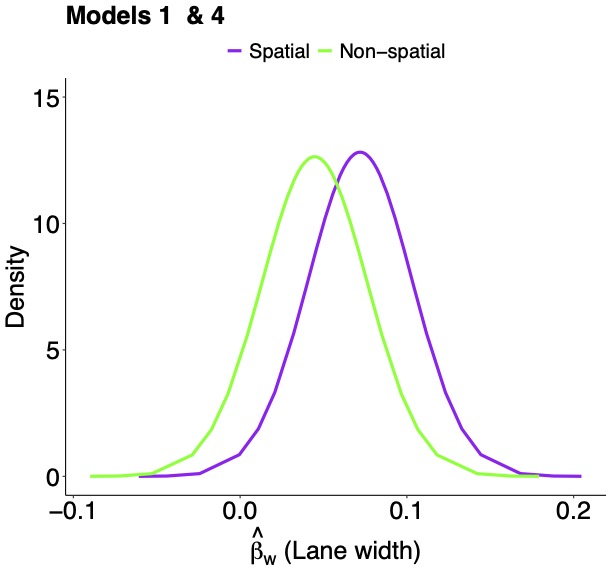}}
	\label{fig:modelwidth}
	\caption{Marginal posterior distributions for rut depth from the previous year and lane width from spatial and non-spatial models.}
	\label{fig:covariatesRut_depthspatialnonspatial}
\end{figure}

Estimates of rut depth from the previous year are smaller, including the uncertainty, from the spatial models compared with non-spatial models (Figure \ref{fig:covariatesRut_depthspatialnonspatial}, graph (a)). Some of this uncertainty of the rut depth from the spatial models is captured in the common spatial field and the random year. Estimated lane width is negative, suggesting that narrower roads allow for more rutting rates; Figure \ref{fig:covariatesRut_depthspatialnonspatial}, graph (b).

\subsection{Inference from random effects}
\label{supplementary:inference_random_effects_supplementary}

The estimated random yearly effects $\hat{\gamma}$ from the non-spatial models are much larger than that of the spatial models. 


\begin{figure}[!ht]
	\centering
	\minipage{0.52\textwidth}
	\includegraphics[width=\linewidth]{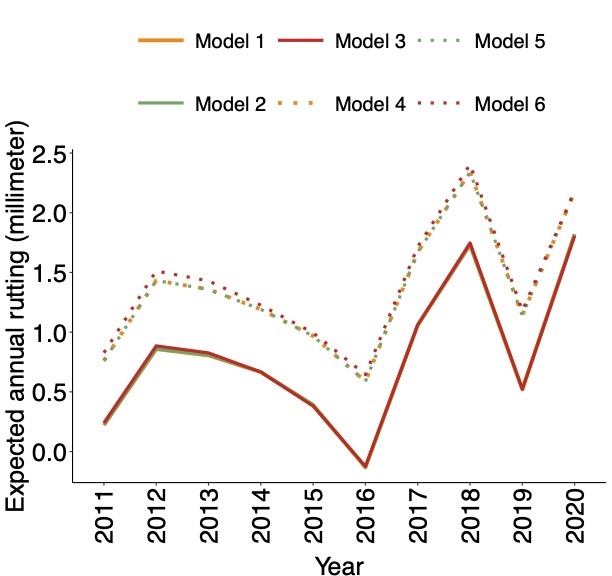}
	\endminipage 
	\caption{Expected annual rutting, $\hat{\gamma}_{t}$. }
	\label{fig:random_year_effect_each_year}
\end{figure}

The estimated nominal range $\hat{\rho}_{t}$ (Table \ref{tab:range_for_each_year_appendix}) is quite variable, especially for the later years 2017-2020. We see that $\hat{\rho}_{2014}$ is also quite large. There were large-scale maintenance activities in 2014 (see  Section ~\ref{supplementary:datasummary}) and, therefore, the correlation between rut depth values will be fast deteriorating with distance, indicating a small spatial range process.  The marginal standard deviation of the spatial field for each year $\hat{\sigma}_{\xi,t}$ (Table \ref{tab:standarddeviation_for_each_year_appendix}) is less than $2.5$ metres, the prior knowledge of the value of GMF at an arbitrary location. This means that with prior knowledge, reasonable and credible intervals are achieved.

\clearpage
\begin{tiny}
	\begin{table}[!htbp]
		\caption{Estimated nominal range for each year $\hat{\rho}_{t}$, with 95\% credible intervals.}
		\label{tab:range_for_each_year_appendix}
		\begin{adjustbox}{width=\textwidth,totalheight=\textheight,keepaspectratio,rotate=0,
				nofloat=table}
			\centering
			\setlength{\tabcolsep}{5.5pt}
			\begin{tabular}{cccccccccccc}
				\hline \\ [1.5ex]
				Year & \multicolumn{2}{c}{\bf Model 1} & \multicolumn{2}{c}{\bf Model 2} & \multicolumn{2}{c}{\bf Model 3}    \\[2.5ex]
				\hline \\[1.2ex]
				& $\hat{\rho}_{t}$ & 95\% CI  & $\hat{\rho}_{t}$ & 95\% CI  & $\hat{\rho}_{t}$ & 95\% CI  \\[1.5ex]
				\hline \\[1.2ex]
				2011 & 97.05   & {[}83.58, 112.00{]}      & 97.09     & {[}83.64, 112.00{]}      & 102.25    & {[}88.45, 117.00{]}      \\[1.2ex]
				2012 & 2108.83  & {[}1512.99, 2900.00{]}  & 2108.60   & {[}1511.44, 2900.00{]}   & 1880.16 & {[}1339.24, 2500.00{]}  \\[1.2ex]
				2013 & 66.81    & {[}53.33, 82.60{]}      & 66.94     & {[}53.43, 82.80{]}       & 73.74    & {[}58.94, 91.00{]}       \\[1.2ex]
				2014 & 3138.64   & {[}1778.63, 5140.00{]} & 3130.05   & {[}1779.154, 5100.00{]}  & 2966.65  & {[}1633.97, 4950.00{]}   \\[1.2ex]
				2015 & 139.62    & {[}108.92, 176.00{]}   & 139.71    & {[}109.02, 176.00{]}     & 139.46   & {[}108.98, 176.00{]}      \\[1.2ex]
				2016 & 562.03  & {[}347.95, 862.00{]}     & 565.73    & {[}349.06 , 873.00{]}    & 537.80   & {[}349.14, 795.00{]}    \\[1.2ex]
				2017 & 5705.58 & {[}3255.58, 9810.00{]}   & 6331.51   & {[}3398.84, 11800.00{]}  & 4286.31  & {[}2743.88, 6420.00{]} \\[1.2ex]
				2018 & 7799.44 & {[}5039.78, 1160.00{]}   & 7911.99   & {[}5162.89, 11900.00{]}  & 6408.31  & {[}2919, 9590.00{]} & \\[1.2ex]
				2019 & 5594.80 & {[}3403.45, 8800.00{]}   & 5572.46   & {[}33865.39, 8790.00{]}  & 4406.75  & {[}2919.47, 6410.00{]}  \\[1.2ex]
				2020 & 8065.79 & {[}5600.78, 11500.00{]}  & 8088.92   & {[}5622.15, 11500.00{]}  & 7246.57  & {[}5121.41, 10200.00{]}  \\[1.2ex]
				\hline \\[0.3ex]
			\end{tabular}
		\end{adjustbox}
	\end{table}
\end{tiny}

\clearpage

\begin{tiny}
	\begin{table}[!htbp]
		\caption{Estimated standard deviation of the spatial field for each year $\hat{\sigma}_{\xi_{t}}$, with 95\% credible intervals.}
		\label{tab:standarddeviation_for_each_year_appendix}
		\begin{adjustbox}{width=\textwidth,totalheight=\textheight,keepaspectratio,rotate=0,
				nofloat=table}
    \tiny
			\centering
			\setlength{\tabcolsep}{5.5pt}
			\begin{tabular}{cccccccccccc}
				\hline \\ [1.0ex]
				Year & \multicolumn{2}{c}{\bf Model 1} & \multicolumn{2}{c}{\bf Model 2} & \multicolumn{2}{c}{\bf Model 3}    \\[2.0ex]
				\hline \\[1.2ex]
				& $\hat{\sigma}_{\xi_{t}}$ & 95\% CI  & $\hat{\sigma}_{{\xi}_{t}}$ & 95\% CI  & $\hat{\sigma}_{{\xi}_{t}}$& 95\% CI \\[1.5ex]
				\hline \\[1.2ex]
				2011 & 1.60  & {[}1.51, 1.69{]}   & 1.60  & {[}1.50, 1.69{]} & 1.62   & {[}1.53, 1.72{]}      \\[1.0ex]
				2012 & 1.31  & {[}1.04, 1.63{]}   & 1.31  & {[}1.04, 1.63{]} & 1.35   & {[}1.09, 1.66{]}     \\[1.0ex]
				2013 & 1.73  & {[}1.62, 1.84{]}   & 1.73  & {[}1.62, 1.84{]} & 1.71   & {[}1.60, 1.82{]}        \\[1.0ex]
				2014 & 1.03  & {[}0.76, 1.38{]}   & 1.03  & {[}0.76, 1.38{]} & 1.04   & {]}0.77, 1.38{]}      \\[1.0ex]
				2015 & 1.08  & {[}0.98, 1.18{]}   & 1.08  & {[}0.98, 1.18{]} & 1.07   & {[}0.98, 1.18{]}       \\[1.0ex]
				2016 & 0.51  & {[}0.42, 0.60{]}   & 0.50  & {[}0.42, 0.60{]} & 0.54   & {[}0.45, 0.63{]}      \\[1.0ex]
				2017 & 1.00  & {[}0.75, 1.31{]}   & 0.99  & {[}0.74, 1.31{]} & 1.02   & {[}0.79, 1.32{]}     \\[1.0ex]
				2018 & 1.39  & {[}1.01, 1.87{]}   & 1.40  & {[}1.02, 1.89{]} & 1.42   & {[}1.06, 1.88{]}    \\[1.0ex]
				2019 & 0.52  & {[}0.37, 0.71{]}   & 0.52  & {[}0.38, 0.79{]} & 0.56   & {[}0.42, 0.75{]}    \\[1.0ex]
				2020 & 1.49  & {[}1.08, 2.02{]}   & 1.49  & {[}1.09, 2.02{]} & 1.51   & {[}1.12, 2.03{]}    \\[1.0ex]
				\hline \\[1.0ex]
			\end{tabular}
		\end{adjustbox}
	\end{table}
\end{tiny}

\vspace{-1.2cm}
\bibliographystyle{tfv}
\bibliography{interacttfvsample}